\documentclass[%
 reprint,
 amsmath,amssymb,
 aps,
floatfix,
]{revtex4-2}

\usepackage{physics}
\usepackage{graphicx}% Include figure files
\usepackage{dcolumn}% Align table columns on decimal point
\usepackage{bm}% bold math
\usepackage[dvipsnames]{xcolor}
\usepackage[danish,spanish,english]{babel}
\begin{document}

\preprint{APS/123-QED}

\title{Violation of Bell’s inequality with quantum-dot single-photon sources}% Force line breaks with \\

\author{Eva M. González-Ruiz}
\email{eva.ruiz@nbi.ku.dk}
\author{Sumanta K. Das}
\author{Peter Lodahl} 
\author{Anders S. Sørensen} 
\affiliation{%
 Center for Hybrid Quantum Networks (Hy-Q), Niels Bohr Institute\\
 University of Copenhagen, Blegdamsvej 17, DK-2100 Copenhagen, Denmark
}%

\date{\today}% It is always \today, today,

\begin{abstract}
We investigate the possibility of realizing a loophole-free violation of Bell's inequality using  deterministic single-photon sources. We provide a detailed analysis of a scheme to achieve such violations over long distances with immediate extensions to device-independent quantum key distribution. We investigate the effect of key experimental imperfections that are unavoidable in real-world single-photon sources including the finite degree of photon indistinguishability, single-photon purity, and the overall source efficiency. We benchmark the performance requirements to state-of-the-art deterministic single-photon sources based on quantum dots in photonic nanostructures and find that experimental realizations appear to be within reach. We also evaluate the requirements for a post-selected version of the protocol, which relaxes the demanding requirements with respect to the source efficiency.
\end{abstract}

\maketitle

%\tableofcontents

\section{\label{sec:intro}Introduction}

Violation of Bell's inequality is of fundamental significance for our understanding of nature and has been the subject of a broad experimental endeavour for the last few decades. A complete violation is highly challenging due to the presence of the locality and detection loopholes \cite{loopholes}, which must be closed simultaneously to realize loophole-free experiments \cite{giustina_2015,Shalm2015,Hensen2015}. While the locality loophole only requires measurement stations to be sufficiently separated, overcoming the detection loophole demands high transmission efficiencies \cite{Pironio_2009}, which is particularly challenging at the large distances required to close the locality loophole. In addition to its fundamental significance, closing the loopholes enables important  technological applications. In particular, a detector-loophole-free violation of Bell's inequality is required for device-independent quantum key distribution (DIQKD) \cite{Pironio_2009}, which allows ultimately secure communication protected even against hacking attempts on the applied hardware. Although these applications do not necessarily need to close the locality loophole, they impose similar demands since cryptography is based on the communicating parties to be far apart.

Recently three different experiments have successfully closed both loopholes simultaneously. On one side, Bell violation has been achieved based on a heralding scheme with NV centers \cite{Hensen2015}, but the operation of these systems is rather complicated and typically slow leading, e.g., to limited key rates for DIQKD. Faster operations have been achieved in purely photonic systems by exploiting efficient superconducting detectors, but at the cost of only working over short distances ($<200$ m) \cite{giustina_2015,Shalm2015}. 

A solution to overcome these problems has been proposed in Ref. \cite{main_article} based on a photonic approach using deterministic single-photon sources. This method exploits heralding measurements at a central station and is thus applicable to long distances with limited transmission probability. If the photons are transmitted through vacuum this could allow the implementation of loophole-free violation of Bell's inequality. For photons transmitted through optical fibers on the hand, the slower propagation speed in the fibers potentially open the locality loophole (discussed in detail below). Nevertheless the proposal still enables closing the detector loophole over long distances and thus the application of the scheme for DIQKD. The scheme could also be implemented with spontaneous parametric down conversion (SPDC) sources, but this was found to have a less favorable  scaling with the transmission efficiency \cite{main_article}. For instance it achieves lowers a key rate in DIQKD and this makes an implementation with on-demand single-photon sources more attractive.
Real single photons, however, have a number of imperfections which could potentially prevent the violation of Bell's inequality. The influence of these imperfections thus needs to be carefully assessed to determine the applicability of existing single-photon sources. 

In this article we theoretically investigate the performance of real single-photon sources for the (detector-)loophole-free Bell test of Ref. \cite{main_article} taking into account the quality of the single-photon sources and the efficiency of the local stations. We focus on InGaAs quantum dots embedded in a photonic-crystal waveguide as single-photon sources. These have recently shown near ideal performance, generating on-demand single photons with up to a 99.4\% purity and 96\% indistinguishability \cite{Uppueabc8268,Tomm:2021vq}. The results derived here \footnote{The codes used in this study are available at the University of Copenhagen public repository ERDA. DOI: https://doi.org/10.17894/ucph.6b93ed79-4a9d-4d34-b5ea-07479288d11d} are also applicable to other types of single-photons sources. We investigate the generic imperfections and analyse the relation between the achievable violation of Bell’s inequality and the indistinguishability of the generated photons through the Hong-Ou-Mandel (HOM) visibility, their second-order correlation function ($g^{(2)}$) and the single-photon source efficiency. We hope that the analysis will motivate experimental demonstrations of long-distance photonics based (detector-)loophole-free Bell inequality violations in the near future.

\section{\label{sec:protocol}Ideal implementation}

\begin{figure*}
\includegraphics[width=\textwidth]{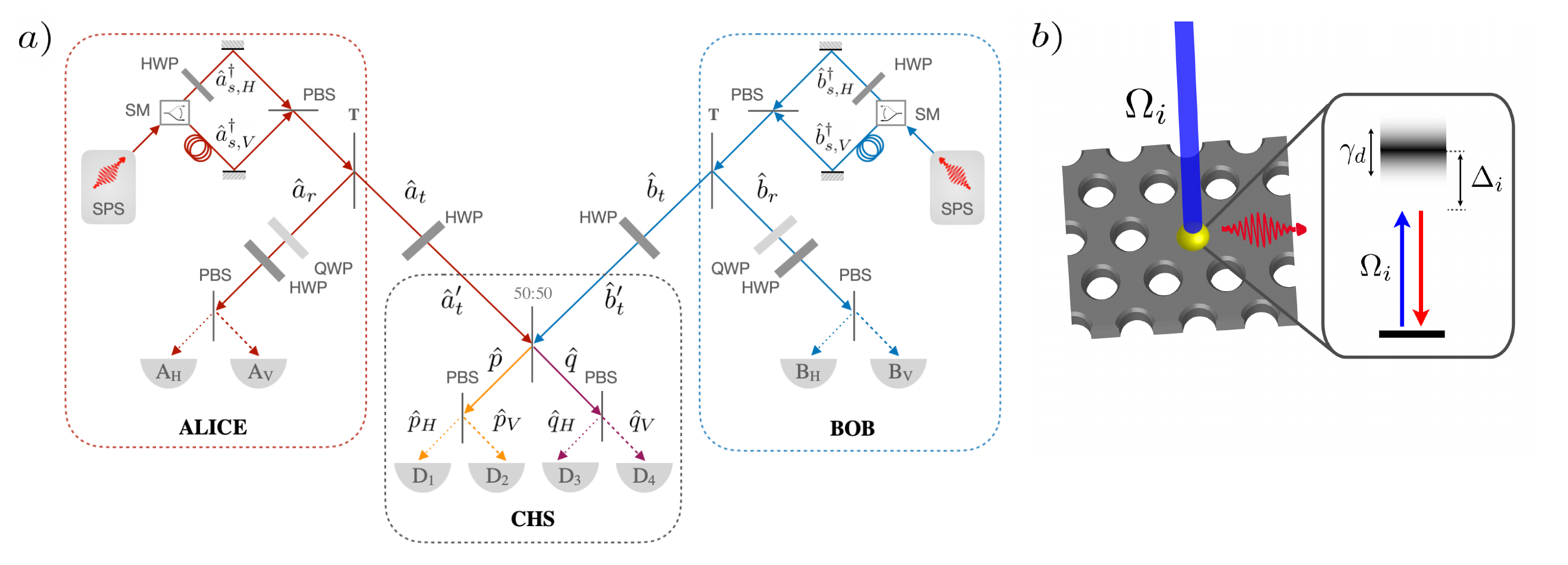}% Here is how to import EPS art
\caption{
\textit{a)} Set-up proposed in Ref. \cite{main_article}, consisting of two local stations (Alice and Bob) and a central heralding station (CHS). The different optical components involved are indicated: beam splitters with transmittances $T$ (at Alice and Bob's stations) and 50\% (at the CHS); half and quarter-waveplates (HWP and QWP), switchable mirrors (SM), polarizing beam splitters (PBS) and single-photon sources (SPS). The photodetectors are labeled as $D_1$, $D_2$, $D_3$, and $D_4$ at the CHS and $A_H$ ($B_H$) and $A_V$ ($B_V$) for Alice's (Bob's) detectors, with the sub index indicating the polarization of the corresponding incoming photons. The operators that model the creation of photons at each section of the set-up are indicated as well. \textit{b)} Example of quantum-dot single-photon source and its energy level schemes. The light-matter interaction between a quantum-dot (yellow dot) and a waveguide mode is enhanced by the nanostructure realizing a deterministic single-photon source. The quantum dot is driven by a resonant excitation pulse $(\Omega_{i})$ and subsequently emits a single photon into the waveguide by spontaneous emission. Insert: The optical transitions are subject to two different types of decoherence: 1) rapidly fluctuating phonon interactions induce pure dephasing with a rate $\gamma_d$ and 2) slow drifts of the levels induce a slowly varying detuning $\Delta_{i}$.}
  \label{fig:CH}
\end{figure*}

We investigate the scheme originally put forward in Ref. \cite{main_article}. It is based on a heralding scheme in which two parties, Alice and Bob, each generate a pair of single photons with orthogonal polarization (see Fig. \ref{fig:CH}). Here we will consider an implementation based on having only a single on-demand single-photon source at each station. As we will show below an implementation based on subsequent emissions from the same source is highly advantageous since several imperfections associated with slow drifts in the sources naturally cancel out.

The single-photon sources each emit a pair of photons with vertical polarization but with one of them delayed in time with respect to the other. The photons are sent to different optical paths by means of a switchable mirror. The early photon is sent through the longest arm, while the late photon is sent to the short arm and is rotated to horizontal polarisation with a half-wave plate. The early and late photons arrive simultaneously at a polarizing beam-splitter (PBS), which merges the two inputs generating a state described by
\begin{align}
\begin{split}
    \hat{a}_{s,H}^{\dagger}\hat{a}_{s,V}^{\dagger}\ket{\emptyset}_A &= \ket{1_H}\ket{1_V}_A, \\ \hat{b}_{s,H}^{\dagger}\hat{b}_{s,V}^{\dagger}\ket{\emptyset}_B &= \ket{1_H}\ket{1_V}_B\,,
\end{split}
\label{eq:initial_mode_operators}
\end{align}
where $ \hat{a}_{s}^{\dagger} $ ($ \hat{b}_{s}^{\dagger} $) denotes the photon creation operator acting on the vacuum state $\ket{\emptyset}$ either at Alice's (A) or Bob's (B) station, with the second index denoting horizontal ($H$) or vertical ($V$) polarization. 

After the PBS, the photon pair is sent towards a beam-splitter of transmittance T. The reflected field is then sent to a detection set-up, whereas the transmitted field is sent towards a central heralding station (CHS) placed between Alice and Bob. At the CHS the photons pass through additional optical components until they are finally detected by four single photon detectors $D_l$ with $l=1,..4$. 
As we will show later, entanglement will be generated if the photons detected at the CHS have orthogonal polarization. This restricts the possible accepted detection patterns to clicks on pairs of detectors $D_1D_2$, $D_3D_4$, $D_1D_4$ and $D_2D_3$ (see Fig. \ref{fig:CH}).

Ideally, two photons (one from each side) are transmitted to the CHS while the other photons remain in the station where they were generated. At the CHS, Bell state measurements on the received photons are performed \cite{BSM} thereby creating a polarization entangled state between the photons that remained in the local stations. Importantly, the protocol is still applicable even for a very low transmission probability to the CHS, since photon loss only reduces the heralding rate but not the performance of the protocol. Finally, Alice and Bob measure their respective qubits by means of a pair of half-wave and quarter-wave plates (HWP and QWP) and a pair of photodetectors.

The basic operational principle of the set up is that of the Hong-Ou-Mandel (HOM) effect \cite{originalHOM}. In each station, beam splitters of transmittance $T$ transmit single photons from Alice and Bob to the CHS. Conditioning on two clicks at the CHS means that at least two photons were transmitted, and by choosing a low transmittance $T\ll 1$ we ensure that there is a negligible probability of transmitting more than two. In this case there will thus be two photons at Alice and Bob's stations for the final Bell's test with a very high probability if $T\ll 1$. The two photons at the CHS photons could, however, be from the same station, such that either Alice or Bob will have two photons while the other would have none. To exclude this possibility a HWP is inserted in each of the arms leading to the CHS. This HWP rotates each of the polarizations by 45 degrees. Seen in the horizontal-vertical basis the HWP essentially acts as a beam-splitter transformation between the two polarization states described by
\begin{equation}
     \hat{a}_{H}^{\dagger} \rightarrow \frac{1}{\sqrt{2}}\left(\hat{a}_{H}^{\dagger} + \hat{a}_{V}^{\dagger} \right), \quad \hat{a}_{V}^{\dagger} \rightarrow \frac{1}{\sqrt{2}}\left(\hat{a}_{H}^{\dagger} - \hat{a}_{V}^{\dagger} \right)\,.
\end{equation}
Thereby it acts as a beam splitter in a HOM-like setup such that two photons in the same arm will bunch and never end up with opposite polarization when measured in the horizontal-vertical basis. Conditioning on photons with different polarization will thus ensure that the detected photons came from different stations and that there is always a photon at both Alice and Bob's stations.

If the two photons transmitted to the heralding station have the same polarization they will bunch at the 50:50 central beam splitter due to their indistinguishability, leading to the possible detector combinations $D_1D_2$ and $D_3D_4$. These patterns do not, however, provide any information about the initial polarization of the photons before the HWP except that they were identical. Thus, the state of the pair that Alice and Bob will share in this case is the Bell state $\ket{\phi^-}=\frac{1}{\sqrt{2}}\left( \hat{a}_H^{\dagger}\hat{b}_H^{\dagger} - \hat{a}_V^{\dagger}\hat{b}_V^{\dagger} \right)\ket{\emptyset}$ (with the minus sign coming from a more detailed analysis of the protocol). 

On the other hand, if the pair of photons arriving at the CHS has opposite polarizations, they are distinguishable and therefore will not show the HOM bunching effect at the beam splitter. If they are transmitted to the same port of the beam splitter they will click on the same detector since the PBS acts as a HOM setup for the photons encoded in diagonal polarizations by the HWPs and will be discarded. On the contrary, if they are transmitted to different PBS, the combinations $D_1D_4$ and $D_2D_3$ can occur and since their polarizations were initially different, the state that Alice and Bob share must be proportional to $\ket{\psi^-}=\frac{1}{\sqrt{2}}\left( \hat{a}_H^{\dagger}\hat{b}_V^{\dagger} - \hat{a}_V^{\dagger}\hat{b}_H^{\dagger} \right)\ket{\emptyset}$ (again the minus sign is found from a more detailed analysis).

From the above analysis we thus find that recording clicks in detectors with opposite polarizations implies that there will be a photon at each of Alice and Bob's stations prepared in an entangled state conditioned on the outcome at the CHS. Subsequent measurements in different bases implemented by rotating the polarization with the HWPs and QWPs at the local stations can then be used to demonstrate the violation of Bell's inequality. The combinations that generate entanglement ($D_1D_2$, $D_3D_4$, $D_1D_4$ and $D_2D_3$) represent 50$\%$ of the total number of detection events. This is in accordance with the fact that any linear-optical circuit that performs Bell-state measurement can only generate distinguishable entangled states with at most 50$\%$ probability \cite{Max50}. In particular, in this work we analyse the state heralded by the detection patterns $D_1D_4$ and $D_2D_3$, which is  $\ket{\psi^{-}}$.

From the above idealized description it is clear that the protocol relies heavily on high-quality single-photon sources. In practice, real sources will always have small multi-photon components and non-perfect indistinguishability. In addition, optical systems are prone to losses. To assess the performance of the protocol for real sources we thus need to evaluate the effect of these imperfections. 

\section{\label{sec:Model}Model for real sources}

The ultimate purpose of this work is to analyse the influence that the quality of a realistic single-photon source can have on the success of the protocol. In this section we present the main parameters that describe the source. We will account both for the loss of indistinguishability of photons due to decoherence processes and the purity of the source, as well as the efficiency of the set-up. 

\subsection{Dephasing}

\begin{figure}
\includegraphics[width=\columnwidth]{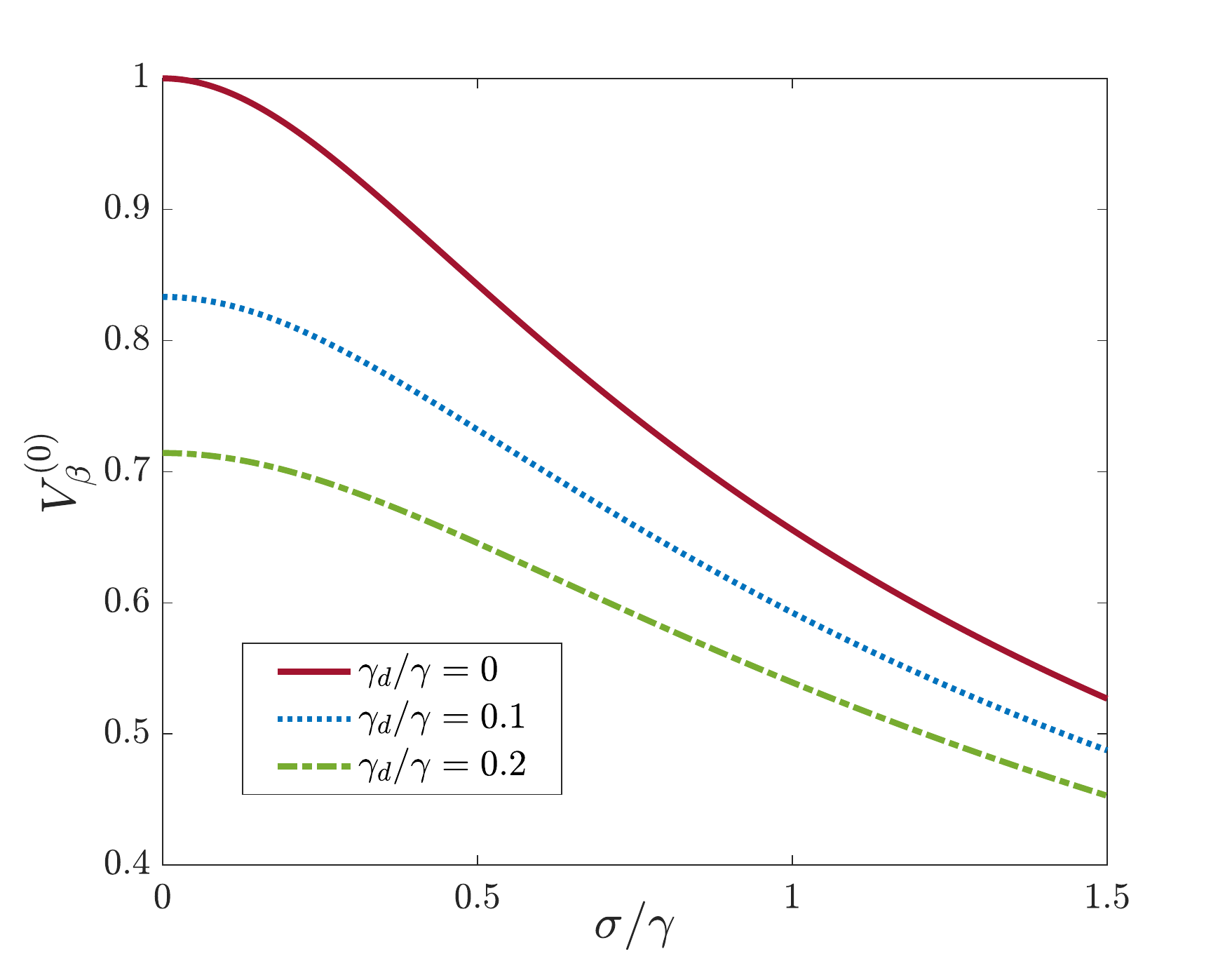}
\caption{Degree of indistinguishability of photons from different sources, i.e. the HOM visibility $V_{\beta}^{(0)}$. The indistinguishability depends on the width $\sigma$ of the probability distribution for the frequency fluctuations $\Delta$ as well as the rate of pure dephasing $\gamma_d$ relative to the spontaneous emission rate $\gamma$.}
  \label{fig:betasq}
\end{figure}

The single-photon source $i$ emits a photon in a state $\hat{a}_i^\dagger\ket{\emptyset}$ . Here the single mode operators $\hat{a}_{i}$ fulfill the single mode commutation relation $[\hat{a}_{i},\hat{a}_{j}^\dagger]=\delta_{ij}$ and are defined by
\begin{equation}
\hat{a}_{i} = \int_{-\infty}^{\infty} f_i(t) \hat{a}_i(t) dt \,,
\label{eq:a_with_f_i}
\end{equation}
where $\hat{a}_i(t)$ is a photon annihilation operator for a photon at time $t$ fulfilling the continuous time commutation relations $\left[\hat{a}(t),\hat{a}^{\dagger}(t')\right] = \delta(t-t'), \quad \left[\hat{a}(t),\hat{a}(t')\right]  = \left[\hat{a}^{\dagger}(t),\hat{a}^{\dagger}(t')\right] = 0 $. We have here introduced the function $f_i(t)$, with $i\in\{1,2,3,4\}$, (different for every source in which the photons are generated) which describes the shape of each photons wavepacket. $f_i(t)$ convolutes each photon's creation operator differently in time, according to the properties of the emitter. 
Ideally $f_i(t)$ will be the same for all sources, resulting in fully indistinguishable photons. Variations between emitters and temporal fluctuations, however, imply that they may vary between different emitters and between different experimental runs.  These functions thus encode the relevant photon coherences. 

The functions $f_i(t)$ satisfy the normalization $\int_{-\infty}^{\infty} |f_i(t)|^2 dt = 1$ and their overlap defines the degree of \textit{indistinguishability} of the photons, that we define as $\alpha_{ij}$ or $\beta_{ij}$ depending on whether the pair of photons was produced at the same or different stations:
\begin{align}
\begin{split}
\expval{\hat{a}_{i}(t)\hat{a}_{j}^{\dagger}(t')} &= \int_{-\infty}^{\infty} f_i^{*}(t) f_j(t) dt \\
&\equiv\begin{cases} \alpha_{ij}, \quad \text{if} \quad ij=12,34 \\ \beta_{ij}, \quad \text{if} \quad ij = 13,14,23,24 \end{cases}\,.
\label{eq:fi_fj}
\end{split}
\end{align}
This distinction between photons generated at the same or opposite stations is important, since their degree of indistinguishability is expected to be different ($\alpha_{ij}>\beta_{ij}$) because sources at the same station are likely easier to be identical. We can relate $\alpha_{ij}$ or $\beta_{ij}$ to the raw HOM visibility $V_{\alpha,\beta}^{(0)}$ by calculating the probability of detecting coincidence counts after a 50:50 beam splitter, obtaining $P_{cc} = \frac{1}{2}\left( 1 - \vert\alpha_{ij}\vert^2 \right)$ (or $\beta_{ij}$, if photons were generated at different stations). Given that the maximum probability of coincidence counts that can be achieved is $P_{max}=1/2$, we obtain the relation between visibility and mode overlap \cite{HOM}:
\begin{equation}
V_{\alpha}^{(0)} = 1 - \frac{P_{cc}}{P_{max}} = \overline{| \alpha_{ij} |^2} \,,
\label{eq:ravHOMv}
\end{equation}
where we have defined $\overline{X}$ to be the average of the quantity $X$ over all experimental runs due to the variability of $f_i(t)$.

The convoluting functions $f_i(t)$ contain the decoherence processes of the quantum dot that decrease the indistinguishability of the emitted photons. Generally decoherence arises due to a number of different mechanisms acting on different time scales. We divide these processes into two categories: slow and fast processes relative to the decay rate of the emitter. For quantum dot sources the fast process originates from phonon dephasing, whereas intrinsic charge or nuclear spin noise as well as drift in the experimental setup will contribute to the slow detuning processes \cite{jacques21}. We assume that the two photons generated at each station are obtained by multiplexing photons emitted from the same quantum dot \cite{thommas}, i.e. photons from the same stations are insensitive to slow processes, which will have the same influence on two consecutive photons, but will be affected by fast processes which change the mode function $f_i(t)$ between the two photons. On the contrary, photons emitted from different stations will also be affected by slow drift depending on the degree to which these can be stabilized between the distant emitters.  

The fast processes are modeled as white noise with a pure dephasing rate $\gamma_d$ corresponding to the typical model used to describe phonon dephasing \cite{Pertru,GoodPetru}. In Appendix \ref{app:dephasing} we show that this leads to a visibility
\begin{equation}
V_{\alpha}^{(0)} = \overline{| \alpha_{ij} |^2} = \frac{\gamma}{\gamma+2\gamma_d}\,,  
\label{eq:alpha_sq}
\end{equation}
where $\gamma$ is the spontaneous decay rate of the quantum dot. 

The crossed visibility, $V_{\beta}^{(0)}$, is also affected by the slow dephasing processes that creates an energy difference $\Delta_{ij}$ in the frequency splitting between quantum dots. This difference is assumed to follow a Gaussian distribution with a width $\sigma$ 
when averaging over the experimental runs. Evaluating the visibility for detuned emitters and performing the average we obtain the averaged visibility
\begin{equation}
V_{\beta}^{(0)}= \overline{\vert{\beta_{ij}}\vert^2} = \sqrt{\frac{\pi}{2}}\frac{\gamma}{\sigma}e^{\frac{(\gamma+2\gamma_d)^2}{2\sigma^2}}\text{erfc}\left(\frac{\gamma+2\gamma_d}{\sqrt{2}\sigma}\right)\,.
\label{eq:beta_sq}
\end{equation}
A similar result was obtained in Ref. \cite{Kambs_2018}.

In Fig. \ref{fig:betasq} we show the relation between visibility and width of the distribution for various dephasing rates. The relations derived in Eqs. (\ref{eq:alpha_sq}) and (\ref{eq:beta_sq}) between the intrinsic parameters of the quantum dot and the indistinguishability of the emitted photons can be used to calibrate the width $\sigma$ to the measured HOM visibility through Eq. (\ref{eq:ravHOMv}). For the remaining of this article all plots will thus be shown as a function of the experimentally accessible visibilities $V_\alpha^{(0)}$ and $V_\beta^{(0)}$ rather than $\gamma_d$ and $\sigma$. We note however, that expressing it in terms of visibilities is dependent on the specific model we have assumed for the noise. Below we shall also need higher order moments e.g. $\expval{\alpha_{ij}\beta_{ik}\beta_{jk}}$. Such higher order correlations cannot be directly related to the visibility, which only depends on second order moments. Any relation between higher order terms and the HOM visibility will thus always be model dependent. 

\subsection{Optical loss}

\begin{figure}
\includegraphics[width=\columnwidth]{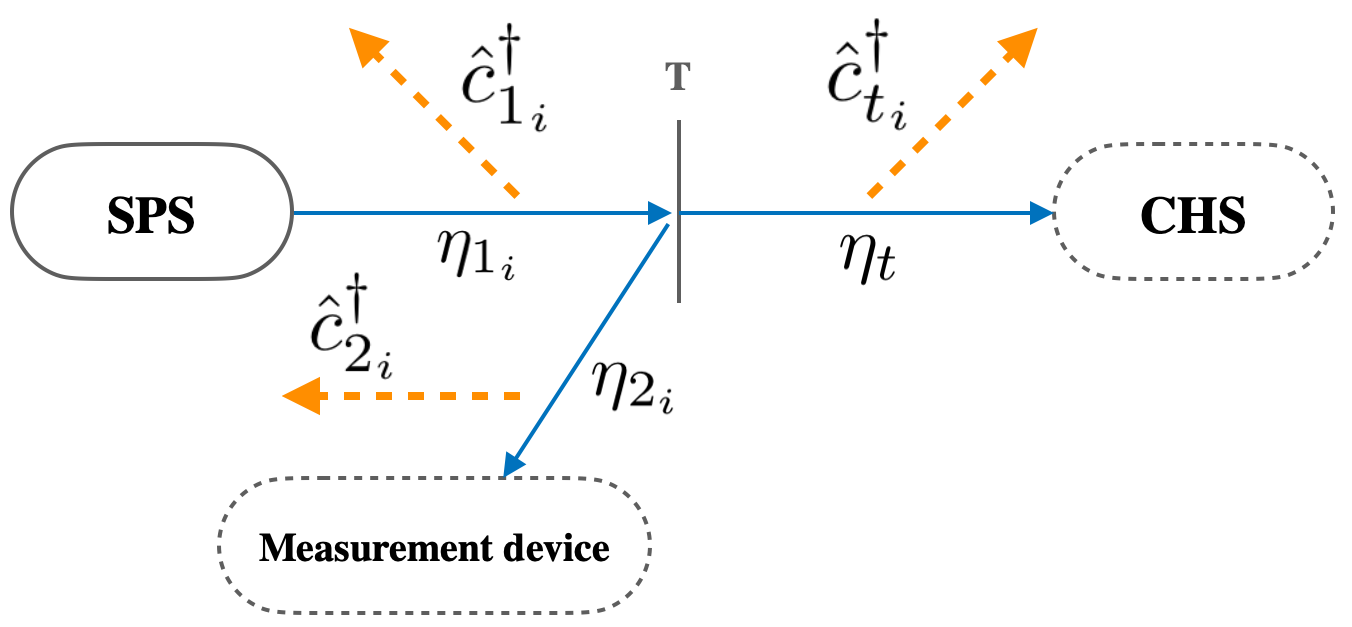}% Here is how to import EPS art
\caption{\label{fig:losses} Sketch of the efficiencies $\eta_{i}$ and definitions of the creation operators of the lost photons $\hat{c}^\dagger$.}
\end{figure}

Although the heralding scheme ensures that the outcome of the protocol is not affected by the transmission efficiency between stations $\eta_t$, we must account for losses occurring locally since such losses potentially open the detection loophole. We define $\eta_{1,i}$ as the probability to reach the first beam splitter, and $\eta_{2,i}$ as the probability for the reflected photons to be detected (including detection efficiency) as sketched in Fig. \ref{fig:losses}. The additional index $i=1,2,3,4$ specifies the source that generated the photon. This distinction between sources is for instance relevant in demultiplexing schemes, which can have different efficiencies in the different channels of transmission \cite{thommas}. Including the losses at the beam splitter, the local efficiency $\eta_{l,i}$ is therefore given by $\eta_{l,i} \equiv \eta_{1,i}\eta_{2,i} (1 - T)$.

Experimentally, the coupling of light from quantum dots to photonic nanostructures has been proven to reach up to 98\% \cite{beta98}. As it will be quantified below, the main challenge will be to couple the photons from the nanostructure and to a detector with near-unity efficiency. With a demultiplexed single-photon source this would require highly-efficient coupling of the photon source to a fiber combined with efficient switching and detection. A significant step towards this demanding goal was the recent demonstration of a fiber-coupled device with $57 \%$ overall efficiency \cite{Tomm:2021vq}. Another implementation would require operating two quantum dots per station that are mutually interfered. In such a configuration the two sources and detectors could potentially be integrated on a single chip, which likely would be the most efficient approach. A bottleneck for this implementation would be the need to make two quantum-dot sources indistinguishable, but  encouragingly $93 \%$ HOM visibility was recently reported between two remote quantum dots \cite{zhai2021}.

\subsection{Purity}

Finally, we consider that the single-photon sources might in fact emit two or more photons at once with a certain probability. This generates the state $\hat{\rho} = P_0 \hat{\rho}_0 + P_1 \hat{\rho}_1 + P_2 \hat{\rho}_2 + \mathcal{O}(P_3)$, where $\hat{\rho}_k$ is the density matrix representing the state of the k-th photon component.  The multi-photon component would originate in an experiment either from imperfect suppression of the optical pulses used to pump the quantum dot or from multi-photon emission due to the finite duration of the excitation pulse.

We furthermore assume that the probability for the quantum dot not to emit a photon is negligible $P_0\simeq 0$, so that $P_1+P_2=1$. This limit is valid as long as the lack of creation of photons by the quantum dot is included in the efficiency parameters $\eta_{1,i}$. 

The effect of multi-photon generation can be measured via the second order correlation function $g^{(2)}$
\begin{equation}
g^{(2)} = \frac{\expval{\iint d\tau_1d\tau_2\hat{a}^{\dagger}(\tau_1) \hat{a}^{\dagger}(\tau_2) \hat{a}(\tau_2) \hat{a}(\tau_1)}}{\expval{\int d\tau_1\hat{a}^{\dagger}(\tau_1)\hat{a}(\tau_1)}\expval{\int d\tau_2\hat{a}^{\dagger}(\tau_2)\hat{a}(\tau_2)}}\,,
\label{eq:g2_def}
\end{equation}
which, for the multi-photon state $\hat{\rho}$ yields
\begin{equation}
    g^{(2)}=\frac{2P_2}{\left( P_1 + 2P_2 \right)^2}\simeq\frac{2P_2}{P_1^2}\,,
    \label{eq:g2_result}
\end{equation}
to first order in $P_2$. Throughout this paper we consider the limit $g^{(2)}\ll 1$ such that the probability that more than one photon is emitted by more than one of the sources is negligible. This means that all results derived here should be considered lowest order expansion in $g^{(2)}$ and are only applicable in the limit $g^{(2)}\lesssim 0.1$. Moreover, the second-emitted photon is modelled to be completely distinguishable from the other four photons. 
The purity of the source will affect the measured HOM visibility $V_{\alpha,\beta}$ with a small contribution proportional to $g^{(2)}$:
\begin{equation}
V_{\alpha,\beta} = V_{\alpha,\beta}^{(0)} - \kappa g^{(2)}\,,
\end{equation}
where in general $1<\kappa<3$, depending on the distinguishability of the 1 and 2-photon components. For the model considered here, $\kappa=2$ where the additional photon is assumed to be fully distinguishable.  \cite{ollivier2020HOM,johannes}. In the following, when plotting quantities as a function of $V_{\alpha,\beta}^{(0)}$ we refer to the value which would be obtained in the absence of any two photon contribution. Hence $V_{\alpha,\beta}^{(0)}$ will be higher than the values measured experimentally. 

\section{Analysis}

Having defined the model used to describe the imperfections, we now turn to the analysis of the performance of the protocol. We do this by analyzing how each of the field operators transform in the Heisenberg picture under the linear optics setup shown in Fig. \ref{fig:CH} when the losses in Fig. \ref{fig:losses} are taken into account. After this transformation we then evaluate all expressions using the model for the real single-photon sources.

After being generated, the two pairs of photons with orthogonal polarization will encounter a beam splitter with transmittance $T$, that transforms the creation operators according to
\begin{align}
\begin{split}
 \hat{a}_{s,{\varsigma}}^{\dagger} &= i\sqrt{1-T}\hat{a}_{\varsigma}^{\dagger}+\sqrt{T}\hat{a}_{t,{\varsigma}}^{\dagger}\,,
\end{split}
\label{eq:BSrelations2}
\end{align}
and similarly for Bob's photons, where $\hat{a}_{\varsigma}^{\dagger}$ represents a reflected photon at the first beam splitter with polarization $\varsigma\in\{H,V\}$ and $\hat{a}_{t,{\varsigma}}^{\dagger}$ a transmitted one. Next, by fixing the angle $\phi = -\pi/8$ for both HWP, we see that the transmitted photons are transformed as:
\begin{align}
\begin{split}
\hat{a}_{s,{\varsigma}}^{\dagger} &= i\sqrt{1-T}\hat{a}_{\varsigma}^{\dagger}+\sqrt{\frac{\eta_t T}{2}}(\hat{a}_{t,H}^{'\dagger}\pm\hat{a}_{t,V}^{'\dagger})\,.
\label{eq:BSrelations3}
\end{split}
\end{align}
where we have inserted the probability of successful transmission between Alice and Bob and the CHS, $\eta_t$. 
Since we include losses, this transformation should, in principle, also include creation operators for the lost photons. However, since we consider photon counting (and multi-photon errors are considered separately below), such lost photons do not contribute to the final results and we omit the lost photon terms for brevity.
Once the photons reach the heralded station they encounter another beam splitter, this time with 50\% transmittance, followed by polarizing beam splitters that direct horizontally (vertically) polarized photons to detectors $D_1$ and $D_3$ ($D_2$ and $D_4$). 

We can now incorporate all the steps into a single transformation of the initial operators $\hat{a}_{S}^{\dagger}$ ($\hat{b}_{S}^{\dagger}$), obtaining
\begin{align}
\begin{split}
&\hat{a}_{{s,\varsigma},{i}}^{\dagger} \rightarrow \\ &\sqrt{\frac{\eta_{1,i} \eta_t T}{2}}\hat{O}_{i}^{\dagger}  + i\sqrt{\eta_{1,i} \eta_{2,i} (1-T)}\hat{a}_{{{\varsigma},{i}}}^{\dagger}+ \hat{L}_{i}^{\dagger}\,,
\end{split}
\label{eq:f_i_operator_relations}
\end{align} 
and similarly for Bob's operators. We have here defined creation operators for photons at the CHS as
\begin{align}
\begin{split}
\hat{O}_{1,2}^{\dagger} &\equiv \frac{1}{\sqrt{2}}\left(i\left(p_{H,{1,2}}^{\dagger}\pm p_{V,{1,2}}^{\dagger}\right)+\left(q_{H,{1,2}}^{\dagger} \pm q_{V,{1,2}}^{\dagger}\right)\right), \\
\hat{O}_{3,4}^{\dagger} &\equiv \frac{1}{\sqrt{2}}\left(\left(p_{H,{3,4}}^{\dagger}\pm p_{V,{3,4}}^{\dagger}\right)+i\left(q_{H,{3,4}}^{\dagger}\pm q_{V,{3,4}}^{\dagger}\right)\right)\,.
\end{split}
\label{eq:O_operators}
\end{align} 
Eq. (\ref{eq:f_i_operator_relations}) expresses that a photon can be either successfully transmitted to the CHS ($\hat{O}_{i}^{\dagger}$), reflected at the first beam splitter and detected locally ($\hat{a}_{{\varsigma},i}^{\dagger}$) or lost ($\hat{L}_{i}^{\dagger}$).
The definition of these operators simplify the transformation applied to the initial state and is described in detail in Appendix \ref{app:BS}.

Eq. (\ref{eq:f_i_operator_relations}) allows us to calculate the density matrix $\hat{\rho}$ shared by Alice and Bob after the heralding by just tracing the CHS and loss operators out from the initial state conditioned on the correct detection patterns:
\begin{equation}
\small
\begin{split}
        \hat{\rho} =  \Tr_{\text{loss},CHS}\left\lbrace\hat{a}_{s,H}^{\dagger}\hat{a}_{s,V}^{\dagger}\hat{b}_{s,H}^{\dagger}\hat{b}_{s,V}^{\dagger}\ket{\emptyset}\bra{\emptyset}\hat{a}_{s,H}\hat{a}_{s,V}\hat{b}_{s,H}\hat{b}_{s,V}\right\rbrace\,,
\end{split}
\end{equation}
where the partial trace of the CHS is already conditioned to the accepted measurement outcomes $D_1D_4$ and $D_2D_3$.

\begin{figure*}
\includegraphics[width=\textwidth]{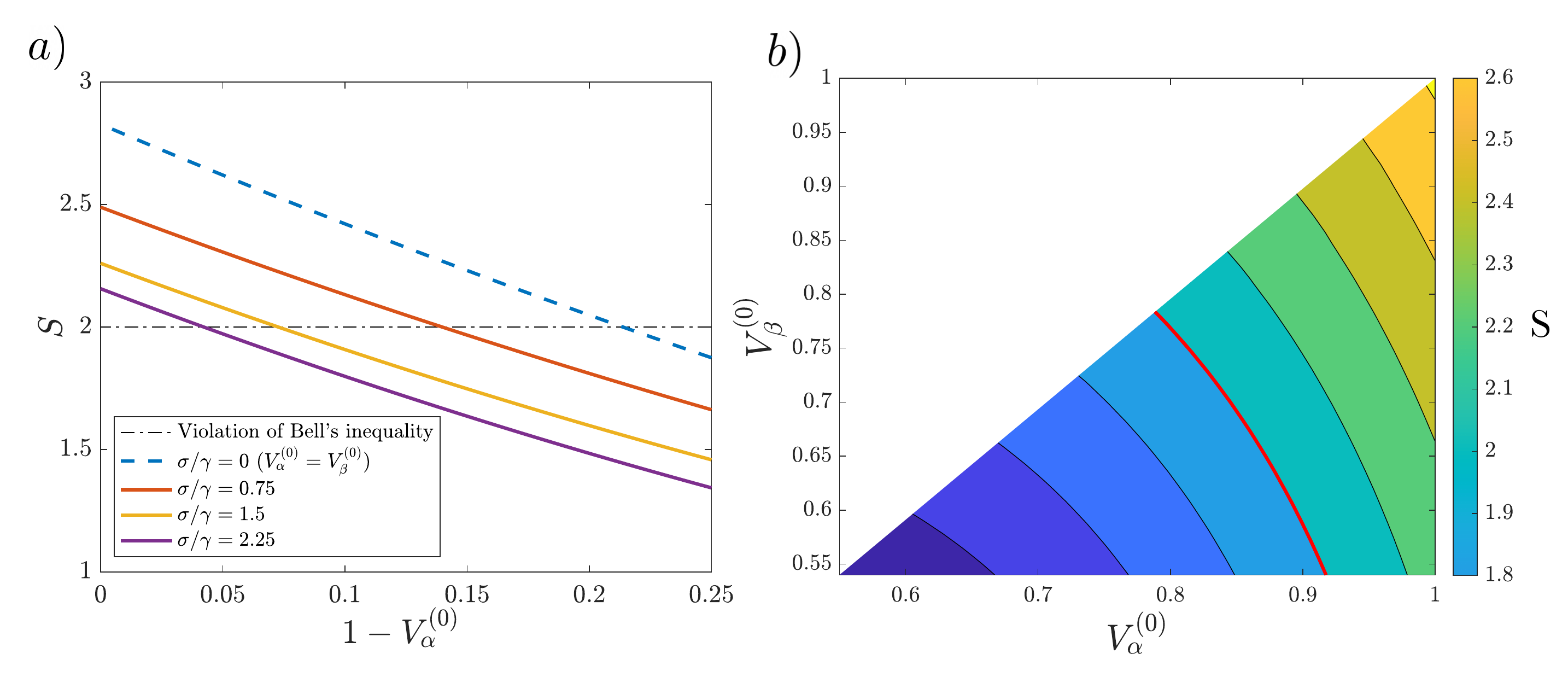}% Here is how to import EPS art
\caption{\textit{a)} Violation of the CHSH inequality $S\geq2$, as a function of the HOM visibility of the local sources $V_{\alpha}^{(0)}$. The curves represent both the case in which photons from Alice and Bob's stations and equally indistinguishable ($V_{\alpha}^{(0)}=V_{\beta}^{(0)}$) and when there is inhomogeneous broadening with a width $\sigma\neq 0$ so that $V_{\alpha}^{(0)}> V_{\beta}^{(0)}$. The threshold for CHSH violation $S=2$ corresponds to the visibilities $(V_{\alpha}^{(0)},V_{\beta}^{(0)}) = \left\lbrace (0.79,0.79),(0.86,0.67),(0.93,0.51),(0.96,0.40)  \right\rbrace$. \textit{b)} Contour plot of the CHSH value $S$ as a function of the visibilities $V_{\alpha}$ and $V_{\beta}$. The red isoline $S=2$ delimits the regions with and without violation of Bell's inequality. }
\label{fig:S_vis_alpha}
\end{figure*}

Alice and Bob measure this state by means of two consecutive QWP and HWP, with the total transformation $\hat{U}_{tot}(\theta,\phi)= \hat{U}_{HWP}(\phi)\hat{U}_{QWP}(\theta)$ (Appendix \ref{app:BS}). It transforms the creation operators of the photons into the operators $\hat{A}_H^{\dagger}$ and $\hat{A}_V^{\dagger}$ ($\hat{B}_H^{\dagger}$ and $\hat{B}_V^{\dagger}$), which represent the creation operators at the corresponding detectors (see Fig. \ref{fig:CH}):
\begin{align}
\begin{split}
\vb{\hat{A}} \equiv \begin{pmatrix} \hat{A}_H \\ \hat{A}_V \end{pmatrix} &= \hat{U}_{tot}(\theta_A,\phi_A)\begin{pmatrix} \hat{a}_{H,{1}} \\ \hat{a}_{V,{2}} \end{pmatrix} \equiv \hat{U}_{tot}(\theta_A,\phi_A) \vb{\hat{a}} \\
\end{split}\,,
\label{eq:a_to_A}
\end{align}
and similarly with Bob's operators. We further define the measurement operators $\hat{M}_{{+},{A}} \equiv \mathbb{I} - \ket{\emptyset}\bra{\emptyset}_{A,{H}}$ and $\hat{M}_{{-},{A}} \equiv \mathbb{I} - \ket{\emptyset}\bra{\emptyset}_{A,{V}}$, and similarly for Bob. These operators project into the subspace in which any number of clicks occurs at the detectors. We thus assume that the detectors cannot distinguish how many photons arrived but only measure the presence or absence of photons. Events in which there is a wrong number of photons involved are thus also included. We identify a click in the detectors $A_H$ and $B_H$ with an outcome $+$, and $A_V$ and $B_V$ with $-$. We can then calculate the joint probability $P_{x,y}$ that the outcomes $x,y \in \left\lbrace +,- \right\rbrace$ are simultaneously measured by Alice and Bob as
\begin{align}
\begin{split}
P_{x,y} = \Tr{\hat{M}_{x,y} \hat{\rho}'}\,,
\end{split}
\label{eq:calculation_of_terms}
\end{align}
where $\hat{M}_{x,y} = \hat{M}_{{x},{A}} \otimes \hat{M}_{{y},{B}}$ and we have defined:
\begin{align}
\begin{split}
\hat{\rho}' = \hat{U}_{tot}(\theta_A,\theta_B,\phi_A,\phi_B) \hat{\rho} \hat{U}_{tot}^{\dagger}(\theta_A,\theta_B,\phi_A,\phi_B)\,,    
\end{split}
\label{eq:rho_prime}
\end{align}
and $\hat{U}_{tot}(\theta_A,\theta_B,\phi_A,\phi_B)=\hat{U}_{tot}(\theta_A,\phi_A)\otimes\hat{U}_{tot}(\theta_B,\phi_B)$. Note that for convenience the probabilities defined here, include the probability of detecting the photons at the CHS. We therefore need to normalize the probabilities with the success probability when we want to evaluate results conditioned on detections at the CHS.
This allows us to calculate the CHSH \cite{CHSH} correlations $C(\textbf{a},\textbf{b})=\textbf{a}\textbf{b}$, where $\textbf{a}$ and $\textbf{b}$ are two unitary observables that can take the values $\{-1,1\}$, as a function of the projected probabilities:
\small
\begin{align}
\begin{split}
&C(\textbf{a},\textbf{b}) = \frac{P_{+,+} - P_{+,-} - P_{-,+} + P_{-,-}}{P_{success}}\,,
\end{split}
\label{eq:final_correlation}
\end{align} 
\normalsize
where the choice of \textbf{a} and \textbf{b} is determined by the HWP and QWP angles $\theta_A$, $\theta_B$, $\phi_A$, and $\phi_B$. Note that the denominator of Eq. (\ref{eq:final_correlation}), $P_{success}$, corresponds to the total success probability for the photons to arrive at the heralding station conditioned on the correct measurement outcomes $D_1D_4$ and $D_2D_3$. We can now analyse how the violation of the CHSH inequality
\begin{equation}
S = | C(\vb{a},\vb{b}) + C(\vb{a'},\vb{b}) + C(\vb{a},\vb{b'}) - C(\vb{a'},\vb{b'}) | \leq 2\,,
\label{eq:CHSHineq}
\end{equation}
is affected by all the processes related to real sources described in Section \ref{sec:Model}.

\subsection{Effect of distinguishability of photons}\label{subsec:indistinsguishability}

We start by analysing solely the influence of decoherence of the emitter, ignoring multi-photon generation ($g^{(2)} = 0$) and local losses ($\eta_{l,i}=1$). Furthermore, we assume for now that apart from the decoherence the protocol behaves ideally: Alice and Bob keep one photon each and send one to the CHS, which is detected in the desired patterns. Physically, this corresponds to post selecting only events where there are always two detected photons at the CHS and one at each station. Applying these conditions to the transformation in Eq. \eqref{eq:f_i_operator_relations} and operating on  the initial state, we obtain:
\begin{align}
\begin{split}
    \ket{\psi} = -&\frac { { \eta_t  }{ T }{ (1-T) } }{ 2 }( \hat{O}_2^{\dagger}\hat{O}_4^{\dagger}\hat{a}_{H,1}^{\dagger}\hat{b}_{H,3}^{\dagger} + \hat{O}_2^{\dagger}\hat{O}_3^{\dagger}\hat{a}_{H,1}^{\dagger}\hat{b}_{V,4}^{\dagger} + \\
    &\hat{O}_1^{\dagger}\hat{O}_4^{\dagger}\hat{a}_{V,2}^{\dagger}\hat{b}_{H,3}^{\dagger} + \hat{O}_1^{\dagger}\hat{O}_3^{\dagger}\hat{a}_{V,2}^{\dagger}\hat{b}_{V,4}^{\dagger})\ket{\emptyset}\,.
\end{split}
\end{align}

We then calculate the partial trace over the the CHS described by the operators $\hat{O}_i^{\dagger}$. Contrary to the ideal case, the state shared by Alice and Bob after conditioning on the desired detection pattern is no longer a pure state due to the imperfect indistinguishability of the photons. 
This can be seen from the (unnormalized) density matrix obtained by conditioning on the correct detection patterns $\expval{\hat{O}_i\hat{O}_j\hat{O}_k^{\dagger}\hat{O}_l^{\dagger}}_{D1D4,D2D3}$. The full results are given in Appendix \ref{app:O}. For simplicity we here only reproduce the density matrix in a simpler form 
assuming $\alpha_{ij}\equiv\alpha$ and $\beta_{ij}\equiv\beta$ for all $i,j$:
\begin{widetext}
\begin{equation}
\small{
 \hat{\rho}  = \frac { { \eta_t  }^{ 2 }{ T }^{ 2 }{ (1-T) }^{ 2 } }{ 4 } \begin{pmatrix} 1- {|\beta|^2} & 0 & 0 & {|\beta|^2}-{|\alpha|^2} \\ 0 & 1 +{|\beta|^2} &  -\left( {|\alpha|^2}+{|\beta|^2}\right) & 0 \\ 0 &  -\left( {|\alpha|^2}+{|\beta|^2}\right) & 1+{|\beta|^2} & 0 \\ {|\beta|^2}- {|\alpha|^2}  & 0 & 0 & 1-{|\beta|^2} \end{pmatrix}}\,,
\label{eq:rho_tr_O_i}
\end{equation}
\end{widetext}
where we have applied the definitions for indistinguishability introduced in Eq. (\ref{eq:fi_fj}). 
The density matrix shown in Eq. \eqref{eq:rho_tr_O_i} is written in a basis that is dependent on polarization, as well as the mode functions which may vary from shot to shot of the experiment. This dependence will be traced out once the measurement at the local stations is performed. 
Note that the state described by Eq. \eqref{eq:rho_tr_O_i} is a mixed state in contrast to the indistinguishable limit ($\alpha = \beta = 1$), in which we in fact recover the pure state $\ket{\psi^{-}}$.

We can now evaluate the results of the measurement with the density matrix $\hat{\rho}$ and calculate the different joint probability contributions by means of Eqs. \eqref{eq:calculation_of_terms} and (\ref{eq:rho_prime}). We note, however, that the density matrix in Eq. \eqref{eq:rho_tr_O_i} is expressed in terms of  photon operators which also contain mode functions that vary in time due to noise. All expectation values will involve a total of 8 mode functions $f_i$, which we need to average over. Therefore, this results in higher order moments in $\alpha_{ij}$ and $\beta_{ij}$, which have to be averaged e.g. giving terms of the form $\overline{\alpha^2\beta^2}$. 

We calculate the value of the CHSH Bell parameter $S$ as a function of the HOM visibility $V_{\alpha}^{(0)}$, that is, the HOM visibility of the sources from the same station (see Fig. \ref{fig:S_vis_alpha}(a)). We study two different limits: first, we consider the limit in which all photons are equally indistinguishable ($\alpha=\beta$). An optimization over the measurement angles is carried out for each value of the visibility in other to maximise $S$. In this limit a HOM visibility of at least $\approx 79\%$ is needed to violate Bell's inequality. We further study the limit in which the cross visibility $V_{\beta}^{(0)}$ from both stations is lower than the visibility from the same station $V_{\alpha}^{(0)}$ by varying the inhomogeneous broadening $\sigma$. As shown in Fig. \ref{fig:S_vis_alpha}(a), with increasing slow spectral diffusion a higher visibility of the photons from the same stations is required to achieve a Bell violation. In Fig. \ref{fig:S_vis_alpha}(b) we show the size of the violation for all sets of visibilities, keeping in mind that $V_\alpha^{(0)} \geq V_\beta^{(0)}$. Note that if a good local visibility $V_\alpha^{(0)}$ is achieved, the requirement for the crossed visibility $V_\beta^{(0)}$ between stations is rather limited. Remarkably and as an exemplary case, for the experimentally realized value of $V_\alpha^{(0)} = 96 \%$ \cite{Uppueabc8268} it suffices to reach $V_\beta^{(0)} > 40 \%$ to violate Bell's inequality. This is an encouraging requirement that seems well within reach with quantum-dot sources.  

\subsection{Losses and multi-photon errors}

\begin{figure*}
\includegraphics[width=\textwidth]{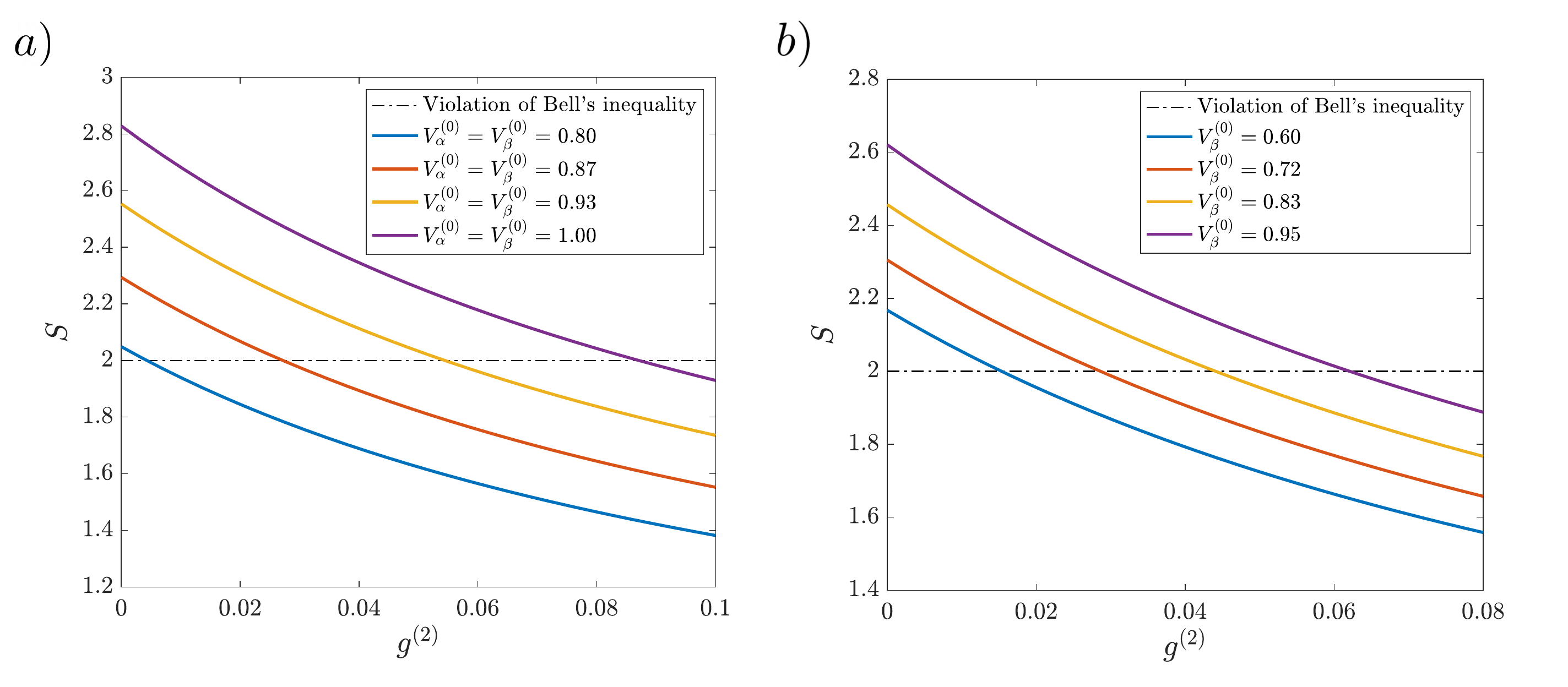}% Here is how to import EPS art
\caption{\label{fig:Sg2_visvis} Evaluation of CHSH threshold for a postselected protocol and with finite multi-photon contributions. In both figures the CHSH parameter $S$ is shown as a function of the second order correlation function for different values of the HOM visibility, in the limit in which the transmission and local efficiencies are low ($\eta_t=\eta_{2,i}=0.1$) and for post-selected events. As post-selection is performed, we set $T=0.5$, which will increase the heralding rate without affecting the performance of the protocol. \textit{a)} The crossed and local HOM visibilities $V_{\alpha}$ and $V_{\beta}$ are identical. \textit{b)} The local HOM visibility is fixed to $V_{\alpha}=0.95$ and $V_{\beta}$ varies between 0.6 and $V_{\alpha}$.}
\end{figure*}

\begin{table*}[htbt]
\centering
\begin{tabular}{rlllllll}
\toprule
 & $P(A=x,B=y)$ & \multicolumn{2}{c}{$P(A=x,B=\emptyset)$} & \multicolumn{2}{c}{$P(A=\emptyset,B=y)$} & \multicolumn{2}{c}{$P(A=\emptyset,B=\emptyset)$} \\ \hline 
4-photon contributions & $P_{2110}$   & $P_{{2200},{SD}}$ &  & $P_{{2020},{SD}}$ &  & $P_{2002}$  & $P_{{2020},{DD}}$ \\
$(i+j+k+l=4)$ &  & $P_{2101}$ & & $P_{2011}$ & & $P_{3001}$  &   \\
 &   & $P_{3100}$ &  & $P_{3010}$  & &  $P_{{2200},{DD}}$  & \\ \hline
5-photon contributions & $P_{2111}$ & $P_{2102}$ & $P_{{3200},{SD}}$ & $P_{2012}$ & $P_{{3020},{DD}}$ & $P_{2003}$ & $P_{{2030},{DD}}$  \\
($i+j+k+l=5$) & $P_{2210,{SD}}$  &  $P_{{2201},{SD}}$ & $P_{2120,{DD}}$ & $P_{{2021},{SD}}$ & $P_{2210,{DD}}$ & $P_{3002}$ & $P_{{3200},{DD}}$\\
 & $P_{2120,{SD}}$  & $P_{{2300},{SD}}$& & $P_{{2030},{SD}}$& & $P_{{2201},{DD}}$ &  $P_{{3020},{SD}}$ \\
 & $P_{3110}$ & $P_{3101}$& & $P_{3011}$ & & $P_{{2021},{DD}}$ & $P_{{2300},{DD}}$ \\
\end{tabular}
\caption{\label{tab:prob_contr_updated} Different contributions to the probabilities $P(A=x,B=y)$, $P(A=x,B=\emptyset)$, $P(A=\emptyset,B=y)$ and $P(A=\emptyset,B=\emptyset)$, where $x$ and $y$ stand for the outcome of Alice and Bob's measurement, respectively. Note that, in the case in which more than one photon is detected at the local stations, we must distinguish between them clicking at the same (denoted with subscript $SD$) or different ($DD$) detectors. 
The first case is seen as an acceptable outcome $x$ or $y$, since we assume that detectors are not number resolving, whereas $DD$ cases are recognized as wrong events by Alice and Bob and are thus assigned to an erroneous outcome $\emptyset$.
}
\end{table*}

When losses and multi-photon errors are included the detection of the correct patterns at the CHS does not guarantee that Alice and Bob share the state predicted by Eq. \eqref{eq:rho_tr_O_i}. For instance, Alice and Bob might get a positive message from the CHS station, but when attempting to measure the state, Bob does not detect anything at his measurement station.
Several events can lead to this situation, depending on whether one of the sources generated more than one photon. Therefore we must identify all of these cases. To simplify the analysis, we introduce the notation $P_{ijkl}$ to denote the probability associated with each of these events. The indices stand for $i$ photons arriving at the CHS, $j$ and $k$ photons being detected by Alice and Bob, respectively, and $l$ photons lost at any point in the set-up. To first order in $g^{(2)}$ and with vacuum emission included in the local efficiency, we can restrict the analysis to total photons numbers of 4 and 5 so that $4\leq i+j+k+l \leq 5$. Moreover, $i\geq 2$, since the accepted detection patterns at the CHS require two photons being detected. Table \ref{tab:events} in Appendix \ref{app:table_probabilities} shows all the different contributions that contribute to each probability term. 

We now divide the contributions into four categories: $P(A=x,B=y)$ with $x,y=\pm 1$ includes all the terms in which both Alice and Bob measure a click in one and only one of their detectors; $P(A=x,B=\emptyset)$ and $P(A=\emptyset,B=y)$ describe events where one of them does not detect any photon and $P(A=\emptyset,B=\emptyset)$ is the probability that none of them detected anything, as specified in Table \ref{tab:prob_contr_updated}. It is important to note that when more than one photon is reflected at the first beam splitter, they can either be detected at the same (denoted with subscript $SD$) or different ($DD$) detectors. In the first situation, this probability will contribute to $P(A=x,B=\emptyset)$ and $P(A=\emptyset,B=y)$, since we assume that detectors are not number resolving and the two photons at the CHS must have originated from the same source. On the other hand, for the $DD$ cases Alice or Bob will know that the result is incorrect. Here we will just assign it to the same category as a lost photon. Furthermore, the two photons from the other station must again have gone to the CHS and these events thus contribute to $P(A=\emptyset,B=\emptyset)$.

We start by considering the effect of multi-photon emission, in the situation typically encountered in current experiments, where the local efficiency is limited and a violation of Bell's inequality can only be obtained by post selection (thereby not closing the detection loophole). In this limit we discard all events in which Alice or Bob did not detect any photon (i.e., the probability terms $P(A=x,B=\emptyset)$, $P(A=\emptyset,B=y)$ and $P(A=\emptyset,B=\emptyset)$). Since this situation is insensitive to local losses it is desirable to set the transmittance to $T=0.5$, which increases the heralding rate. Therefore, compared to the situation in Sec. \ref{subsec:indistinsguishability}, the only new contributions to the CHSH correlations are the 5-photon terms $P_{2111}$, $P_{2210}$, $P_{2120}$ and $P_{3110}$ (see Table \ref{tab:events}).

As mentioned above we will only go to lowest order in the two photon emission probability $P_2$ (or equivalently $g^{(2)}$ according to Eq. \eqref{eq:g2_result}). We can thus restrict the analysis to events where there are 4 and 5 photons in total and the final probability distributions are therefore given by a new probability distribution for the accepted events
\begin{align}
\begin{split}
    P_T&(A=x,B=y) = \\ &\frac{P_1^4P_4(A=x,B=y)+P_1^3P_2P_5(A=x,B=y)}{P_1^4 + 4P_2P_1^3}\,,
\end{split}
\label{eq:g2_P_1}
\end{align}
where $P_4(A=x,B=y)$ and $P_5(A=x,B=y)$ correspond to the probability of 4 and 5 photon event contributions, respectively. Here $P_5(A=x,B=y)$ accounts for the possibility of any of the four single-photon sources emitting two photons. Applying Eq. \eqref{eq:g2_result} we obtain that the total probability that accounts for all the events equals
\begin{align}
\begin{split}
    P_T&(A=x,B=y) = \\ &\frac{P_4(A=x,B=y)+\frac{1}{2}g^{(2)}P_5(A=x,B=y)}{1+2g^{(2)}}\,.
\end{split}
\label{eq:g2_P_2}
\end{align}
\vspace{2mm}
This allows us to calculate the CHSH S-parameter by means of Eq. \eqref{eq:final_correlation} as a function of the second order correlation function $g^{(2)}$, the efficiencies of every channel, and the indistinguishability of the photons through the moments of $\alpha_{ij}$ and $\beta_{ij}$.

The value of the CHSH parameter $S$ is plotted as a function of the second order correlation function $g^{(2)}$ in Fig. \ref{fig:Sg2_visvis}. We observe a Bell violation for low values of $g^{(2)}$ that ceases once the two-photon probability exceeds a certain value of not more than 10 \%. The robustness towards multi-photon contributions is dependent on the HOM visibility of the photons. The predicted values are compatible with state-of-the-art single-photon sources, where $g^{(2)} \lesssim 2\%$ is currently achieved by quantum-dot single-photon sources embedded in photonic nanostructures \cite{QD5,Tomm:2021vq,Uppueabc8268}. In the figure we focus on the regime corresponding to a typical post-selected experiment where the local and transmission efficiencies are rather small $\eta_t=\eta_{2,1}=0.1$. These presented results thus roughly represent the limit $\eta\rightarrow 0$. For the situations considered below with higher higher local efficiency $\eta_{2,i}\approx 1$ we find that the scheme is more robust to  $g^{(2)}$.

\begin{figure}
\includegraphics[width=\columnwidth]{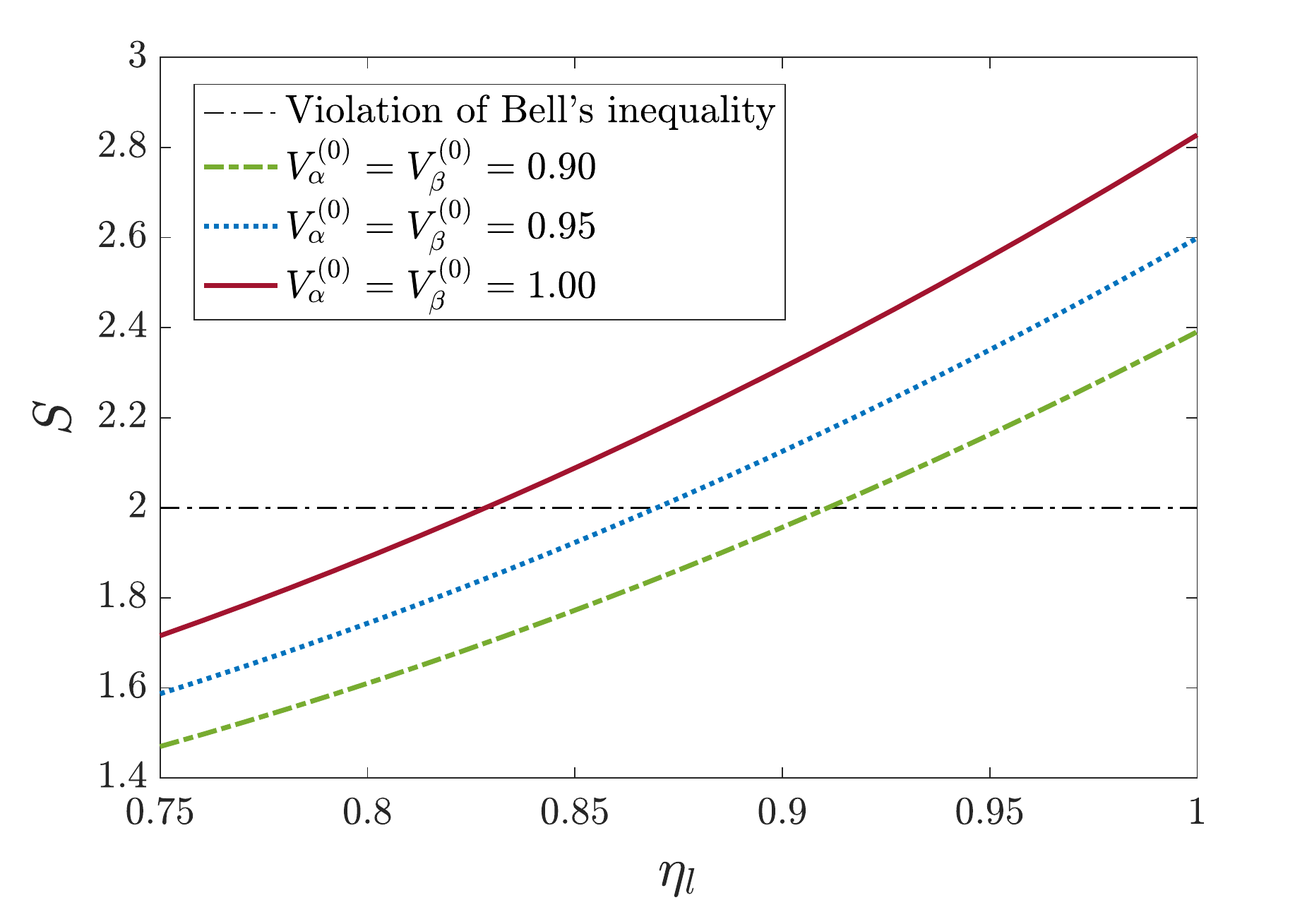}% Here is how to import EPS art
\caption{\label{fig:S_etaloc} CHSH parameter $S$ as a function of the local efficiency $\eta_l=\eta_{1,i}\eta_{2,i}(1-T)$ of Alice and Bob's stations for different HOM visibilities. To focus on the effect of the local efficiency we consider a situation with a very limited transmission to the CHS ($T=10^{-3}$ and $\eta_t=0.1$) and vary the efficiency of the final arm $\eta_{2}$ with $\eta_{1,i}=1$.}
\end{figure}
 
Finally we consider the detection loophole by accepting all events (including those in which no photons are detected, i.e. $P(A=\emptyset,B=y)$, $P(A=x,B=\emptyset)$, $P(A=\emptyset,B=\emptyset)$). In the derivation of the CHSH inequality it is assumed that the measurement can only take on two values -1 and 1. Experimentally we will also have the possibility of an inconclusive outcome $\emptyset$, so we must decide on a strategy for dealing with those outcomes. 
Here we chose that every time Alice and Bob do not detect a photon they assign a determined outcome to it \cite{Pironio_2009} such that there are only two possible outcomes of the experiment that we can assign to $\pm1$, as required by the CHSH inequality. In particular, Alice and Bob assign a positive detection $x,y=\lbrace+\rbrace$ whenever they do not detect any photon. Thus, we arrive at a probability $\tilde{P}$ distribution
\begin{align}
\begin{split}
&\tilde{P}(A=x,B=y) = \\
&P(A=x,B=y) + \delta_{y,+}P(A=x,B=\emptyset) \\ &+ \delta_{x,+}P(A=\emptyset,B=y) + \delta_{x,+}\delta_{y,+} P(A=\emptyset,B=\emptyset)\,.
\end{split}
\label{eq:strategy1}
\end{align}
This allows us to study the effect of the local efficiency on the performance of the protocol (see Fig. \ref{fig:S_etaloc}) independently of the other effects, by setting $g^{(2)}=0$, choosing different values of HOM visibility, and calculating the effective probabilities by applying Eq. \eqref{eq:strategy1} to Eqs. \eqref{eq:g2_P_1} and \eqref{eq:g2_P_2}.

\begin{figure}
\includegraphics[width=\columnwidth]{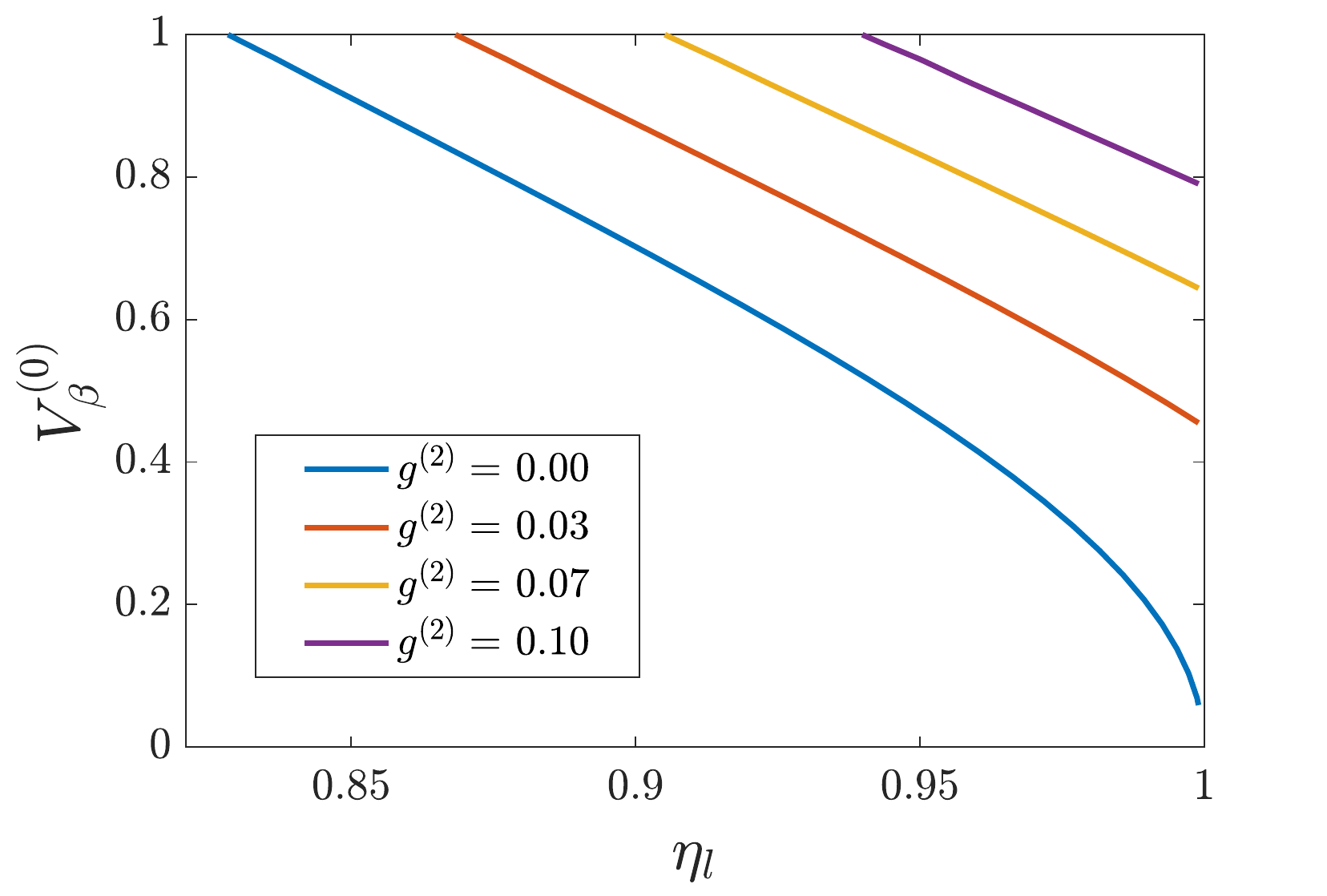}% Here is how to import EPS art
\caption{\label{fig:Vbeta_etaloc} Threshold for the violation of the CHSH Bell's inequality $(S\leq 2)$. 
We vary $\eta_{2,i}$ for different values of the second order correlation function and determine the threshold visibility $V_\beta^{(0)}$ required to violate the CHSH inequality. The local HOM visibility has been set to $V_{\alpha}^{(0)}=1$ and we assume a low transmission efficiency ($T=10^{-3}$, $\eta_t=0.1$).}
\end{figure}

To mimic a situation corresponding to long distance communication where there is limited transmittance and a negligible probability to detect more than one photon at the CHS, we consider a very low transmittance of the beam splitter, $T=10^{-3}$, and a transmittance to the CHS $\eta_t=0.1$. We vary the local efficiency by changing the loss rate in the detection arm $\eta_2$ with $\eta_1=1$. These choices, however, have very little influence on the results as long as we can neglect multiphoton events at the CHS. For perfect HOM visibility, the threshold that delimits $S=2$ is $\eta_l=82.8\%$, which corresponds to the lowest efficiency that any CHSH Bell test can tolerate for loophole-free violation \cite{eta8284}. The simple strategy for dealing with null detections, as introduced above, thus have similar performance as the best achievable strategy. As we introduce additional errors, see Fig. \ref{fig:Vbeta_etaloc}, the required local efficiency increases and goes beyond 90\% for realistic values of $g^{(2)}$ and HOM visibility. Such high local efficiency is not yet  achievable in state-of-the-art implementations, although a lot of progress in this direction has been achieved in recent years \cite{Tomm:2021vq,Uppueabc8268}. Note that compared to the post-selected limit (Fig. \ref{fig:Sg2_visvis}), the violation of the inequality is less sensitive to $g^{(2)}$. This is due to the difference in the local efficiency in the two plots. In the post selected limit events in which two photons are emitted have a higher chance of being accepted relative to the single-photon events, because either of the two photons can make it to the detectors. Had we considered post selection in the limit of high local efficiency, it would be less sensitive to  $g^{(2)}$ since some undesired results, e.g. $DD$ events, can be discarded.

\subsection{Optimizing the probability of transmission $T$}

In previous sections we have calculated and optimised the CHSH value $S$ within the limit of a very small transmission to the CHS. This ensures a higher local efficiency and thus allows us to investigate how well the protocol might ideally work as well as identify thresholds for the success of the protocol.
A low probability of transmission, however, also implies that the number of successful heralding events will be low. In this subsection we find the optimal transmittance $T$ of the beam splitters at the local stations (see Fig. \ref{fig:CH}(a)) to violate Bell's inequality with the highest number of standard deviations $\sigma_S$. 

\begin{figure*}
\includegraphics[width=\textwidth]{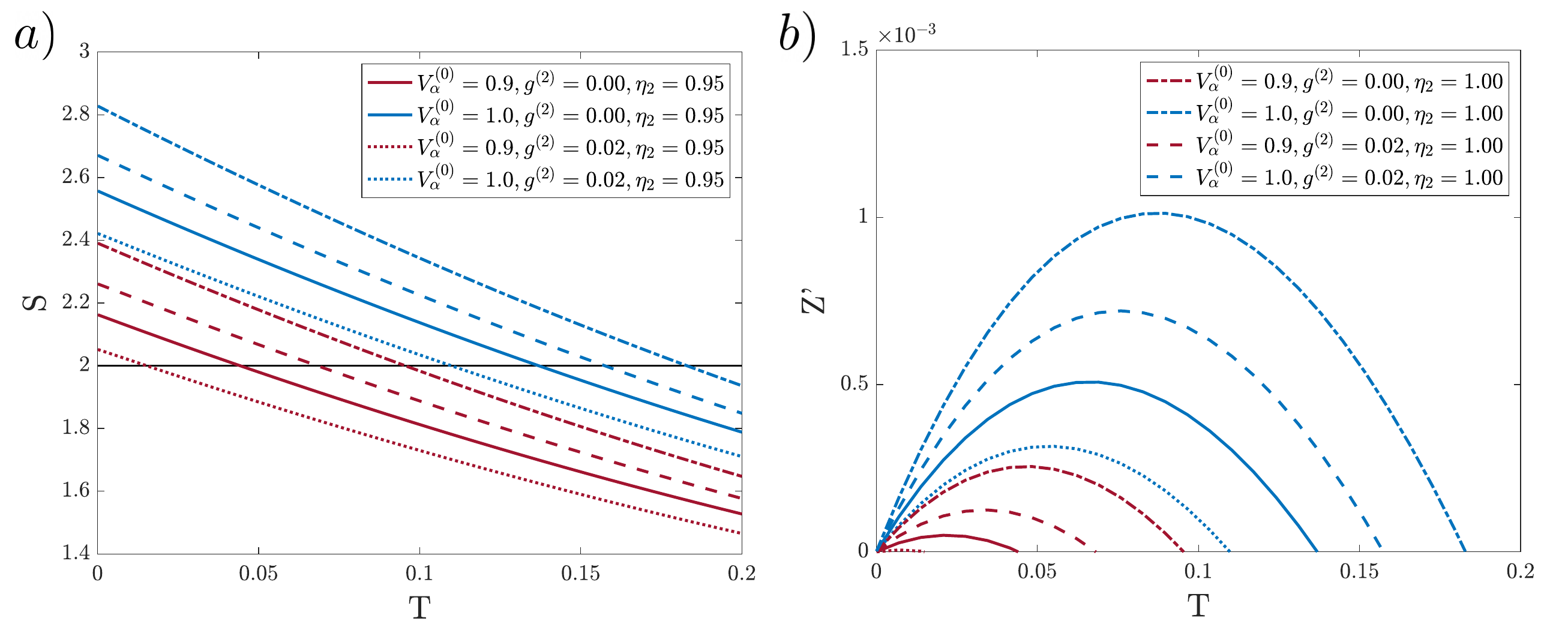}% Here is how to import EPS art
\caption{\label{fig:Toptim} Effect of finite transmittance $T$. \textit{a)} CHSH parameter $S$ as a function of the transmittance $T$ of the beam-splitters at the end stations for several sets of single-source error parameters: visibility ($V_{\alpha}^{(0)}=V_{\beta}^{(0)}$), purity ($g^{(2)}$), and the local efficiency ($\eta_2$ with $\eta_{1,i}=1$). A large transmission probability decreases the performance of the protocol since there will be more events where no photons are detected at end stations.
\textit{b)} Number of standard deviations $Z'$ with which a violation of Bell's inequality is achieved (normalised to the number of experimental runs) as a function of the transmittance $T$. In an experiment with $n$ experimental runs the CHSH inequality is violated by $Z'\sqrt{n}$ standard deviations. An optimal $T$ value that maximises the Bell violation can be found for each parameter set. 
Note that the legend is divided into two parts inserted in part \textit{a)} and \textit{b)} but applies to both subfigures \textit{a)} and \textit{b)}, i.e. half the lines are defined by the legend in \textit{a)}, the others in \textit{b)}.
%Note that the legends apply to both subfigures but are divided between \textit{a)} and \textit{b)}. Both subfigures are plotted in the limit of low transmission to the CHS ($\eta_t=0.1$).
}
\end{figure*}

The CHSH parameter (Eq. (\ref{eq:CHSHineq})) is the sum of four independently measured correlations $C(\vb{a},\vb{b})$. This allows us to write the standard deviation $\sigma_S$ of the CHSH parameter $S$ as
\begin{align}
\begin{split}
\sigma_S^2 &= \expval{S^2}-\expval{S}^2   \\
&= \sigma_{C(\vb{a},\vb{b})}^2 + \sigma_{C(\vb{a}',\vb{b})}^2 + \sigma_{C(\vb{a},\vb{b}')}^2 + \sigma_{C(\vb{a}',\vb{b}')}^2\,.
\end{split}
\label{eq:sigma1}
\end{align}
Furthermore, given that $ C^2(\vb{a},\vb{b}) = \vb{a}^2\vb{b}^2 = 1$ we obtain the standard deviation of $S$ for a single run of the experiment
\begin{align}
\begin{split}
\sigma_S^2 = 4 - C(\vb{a},\vb{b})^2 -C(\vb{a}',\vb{b})^2 - C(\vb{a},\vb{b}')^2 -C(\vb{a}',\vb{b}')^2\,.
\end{split}
\label{eq:sigma4}
\end{align}

After a number of independent experimental runs $n$, there have been $N=n P_{CHS}$ successful events, where $P_{CHS}$ is the total probability of acceptance, equal to the sum of all the probability contributions that return a positive message from the CHS. The standard deviation of the average of $S$ yields
\begin{equation}
\bar{\sigma}_S = \frac{\sigma_S}{\sqrt{N/4}}= \frac{2\sigma_S}{\sqrt{n P_{CHS}}}\,,
\label{eq:nCHS}
\end{equation}
where the factor of 4 arises from the four different measurement configurations of the CHSH parameter $S$. We define $Z$ as the number of standard deviations with which the Bell inequality can be violated
\begin{equation}
Z = \frac{S-2}{\bar{\sigma}_S}=\frac{S-2}{2\sigma_S}\sqrt{n P_{CHS}}\,,
\label{eq:Z1}
\end{equation}
where we have inserted Eq. (\ref{eq:nCHS}) and take into account that the violation occurs for $S>2$.

We have optimized the set of angles for the final measurement at each station for various sets of visibility, losses and multi-photon parameters and a variable transmittance T. Using Eq. (\ref{eq:Z1}) this allows us to calculate a dimensionless number $Z'\equiv Z/\sqrt{n}$ expressing the number of standard deviations normalised by the number of experimental runs. The number of standard deviations obtained after $n$ experimental runs is then $Z'\sqrt{n}$. As we can observe in Fig. \ref{fig:Toptim}(b), the compromise between a strong violation and the probability of success manifests itself in different optimal transmission coefficients that maximises $Z'$ for different parameter sets.
Generally the optimal transmission decreases as other errors become more significant, since in that case there is less room for errors introduced by the transmission. 
Furthermore, in the limit of low transmission efficiency ($\eta_t = 0.1$), the main contributions to the probability of acceptance at the CHS arise from events in which two photons arrive (and not three). This leads to $P_{CHS}\propto T^2$ resulting in a linear increase of $Z'\propto \sqrt{P_{CHS}}\propto T$ for small $T$, as observed in Fig. \ref{fig:Toptim}(b).

Having calculated the optimal $T$, it is straightforward to evaluate the number of experimental runs $n$ necessary to violate Bell's inequality with a certain number of standard deviations $Z$ 
\begin{equation}
n = \left(\frac{Z}{Z'}\right)^2\,.
\label{eq:std_dev}
\end{equation}
Current experiments can easily reach $V_{\alpha}^{(0)} \geq 0.9$ and $g^{(2)}=0.02$ \cite{Tomm:2021vq,Uppueabc8268} 
and for simplicity we assume that a similar visibility is reached between different stations $V_{\beta}^{(0)}=V_{\alpha}^{(0)}$.
A much more challenging requirement is the local efficiency. If we assume optimistic values of $\eta_1 = 1$
and $\eta_2 = 0.95$, we get $Z'=6.2\cdot10^{-6}$ with $\eta_t=0.1$ at the optimal transmission. This means that Bell's inequality can be violated with three standard deviations after $2.3\cdot10^{11}$ runs. Note that for the sake of simplicity, we use  the number of standard deviations with which a Bell violation is achieved as the figure of merit. For a more precise characterization of the violation in an actual experiment it would be desirable to consider the the p-value  \cite{p_value}. Considering the number of standard deviations, however, allows for a simpler evaluation of the requirement to achieve a violation by a different amount $Z$ by means of Eq. \eqref{eq:std_dev}.

For a typical repetition rate of the laser that excites the quantum-dot single-photon sources ($\simeq$ 75 MHz \cite{Uppueabc8268,Tomm:2021vq,Martin1}) this approximately corresponds to only 52 minutes. If we further assume an attenuation length of 20 km, given that the data was obtained for $\eta_t=0.1$, this allows for a separation distance between Alice and Bob's stations of more than 90 km. This is a promising result for future experimental implementations. 

\subsection{Memory considerations}

For experiments based on photons emitted from a central source the locality loophole can typically be closed by rapidly switching the measurement basis at each station. The requirement for closing this loophole is then achieved by having the measurement stations sufficiently separated such that no signalling is possible, i.e. by placing the detectors outside their respective light cones.

Note, however, that for protocols based on heralding this requirement concerns the heralding station as well as Alice and Bob's stations. For photons transmitted through vacuum this can be achieved by adding only a minor delay at each station corresponding to the time it takes to detect the photons at the CHS. On the other hand if Alice and Bob are far apart it is desirable to use optical fibers for the transmission of photons between the measurement stations and the CHS. Since the speed of light is reduced when travelling in the fiber, the heralding is likely to happen inside the light cone of the stations, thus opening the locality loophole. 

This problem can be solved by using local memories at the measurement stations of Alice and Bob. They would then need to store their corresponding photons until the heralding is outside the light cone. In this way, any signalling produced by the CHS would not affect the measurement of the state of the photon, thus closing the locality loophole. State-of-the-art quantum memories have reached an efficiency above 85\% \cite{memory_ref} which is in principle sufficient to allow an efficient violation using this setup, but this would put very stringent requirements on the other local efficiencies. On the other hand for the application of the considered setup for DIQKD, the locality loophole is less important and the main challenge is to close the detection loophole, which can still be achieved without memories. 

\section{Conclusion and outlook}

We have studied the feasibility of using single-photon sources for loophole-free Bell test following the proposal of Ref. \cite{main_article}. The method is general and applies to any single-photon sources, but for concreteness we focus on sources based on quantum dots. Since the considered protocol is completely photonic and has a built-in robustness to losses in the transmission to the CHS, it is highly promising for violating Bell inequalities over long distances, although the reduced transmission speed of optical fibers will likely open the locality loophole for fiber based  implementations. Beyond their fundamental interest such detection-loophole-free violations are of immense technological interest since they allow DIQKD, providing ultimate security in communication. 

The success of the protocol is strongly dependent on the visibility, purity, and efficiency of the
single-photon sources. We find that a HOM visibility of at least 79\% and second-order correlation function $g^{(2)} < 10\%$ suffice for violating the CHSH inequality when post-selecting events (not loophole-free). Such metrics are already demonstrated with state-of-the-art quantum dot sources \cite{Tomm:2021vq,Uppueabc8268}. 
Performing a fully loophole-free violation of Bell's inequality, however, remains a challenge since this puts very stringent requirements on the efficiency of the single-photon sources. For realistic parameters we find that efficiencies on the order of $\eta = 90\%$ are required. This is beyond what has been achieved so far, but continuous improvements in efficiencies have been achieved in recent years. We thus believe that sources capable of achieving such efficiencies will be available in the future. Once the necessary efficiencies are achieved it is highly encouraging that we find the requirements for the crossed visibility between stations to be lower than those for photons generated at the same station. For instance comparing Figs. \ref{fig:betasq} and \ref{fig:Vbeta_etaloc} we see that if the local indistinguishability is high ($V_\alpha^{(0)}\approx 1$) we may obtain a violation even if the spectral fluctuations between the two quantum dots are comparable to the decay rate $\sigma \sim \gamma$. This highly relaxed performance reduces considerably the experimental complexity.

The stringent requirements on the efficiencies may be remedied by designing more advanced protocols. In particular it has been shown that by using non-maximally entangled states and Eberhard's inequality (or, equivalently, CHSH with the assignation strategy used in the present work) the efficiency threshold for Bell tests can be lowered to $\eta=66.7\%$ \cite{eberhard}. For the current setup such non-maximally entangled states can by obtained by replacing the beam-splitter at the CHS by a beam-splitter with a different transmission \cite{main_article}. It is therefore very likely that a similar advantage can be gained for this system. It has, however, recently been proven that the security of some generalised DIQKD protocols involving non-maximally entangled states can be bounded to a similar threshold as the CHSH inequality \cite{CHSHAcin,Sekatski2021deviceindependent}. It is thus unclear if a similar advantage can also be obtained for DIQKD. 
In the future it will be highly interesting to explore the precise relation between the results we have obtained here and the conditions for performing DIQKD as well as the possible advantages of changing to non-maximally entangled states.

\vspace{2mm}

\begin{acknowledgments}
We would like to thank Matthias Christandl for fruitful discussions. We acknowledge the support of Danmarks Grundforskningsfond (DNRF 139, Hy-Q Center for Hybrid Quantum Networks).
\end{acknowledgments}

\onecolumngrid
\appendix

\section{Dephasing in single-photon emitters}\label{app:dephasing}

To evaluate the performance of the protocol we need to evaluate the indistinguishability parameters $\alpha_{ij}$ and $\beta_{ij}$ (and higher order moments). In this section we calculate these for a particular model of the dephasing, which was also considered in \cite{Kambs_2018}. 
We consider a Hamiltonian that describes the dynamics of the system, containing the field, the emitter and their interaction \cite{sumanta}:
\begin{equation}
    \hat{H} = \left(\omega_{eg}^{i} + \Delta_i + F_i(t) \right)\ket{e_i}\bra{e_i} + \int{dk \omega_k \hat{a}_k^{\dagger}\hat{a}_k}  -\int{ dk g_k\ket{e_i}\bra{g_i}\hat{a}_ke^{ikz_i} + h.c.}\,,
\end{equation}
where $\omega_{eg}^{i}$ is the natural frequency splitting of each emitter, perturbed by a slow frequency drift $\Delta_i$ and rapidly varying (e.g. phonon-induced) random fluctuations represented by the uncorrelated force $F_i(t)$. Solving the time evolution of the system through Schr\"odinger's equation with a suitable wavefunction ansatz \cite{sumanta}
allows us to find the mode functions $f_i(t)$ appearing in Eq. \eqref{eq:a_with_f_i}:
\begin{align}
f_i(t,t_0) = \sqrt{\gamma}e^{-\frac{\gamma}{2}(t-t_0)}e^{-i\left(\Delta_i(t-t_0) + \int_{t_0}^{t} F_i(t')dt'\right)} \equiv \sqrt{\gamma}e^{-\frac{\gamma}{2}(t-t_0)}e^{-i\left(\Delta_i(t-t_0) + \phi_i(t,t_0)\right)}\,,
\end{align}
where we account for the spontaneous decay (with spontaneous emission rate $\gamma$) and dephasing of the emitter through the random phase $\phi_i$. We assume that the uncorrelated force $F_i(t)$ satisfies $\expval{F_i(t)}=0$ and \cite{bylander}
\begin{equation}
\expval{F_i(t)F_j(t')} = 2\gamma_d \delta_{ij} \delta(t-t')\,,
\label{eq:languevin_correlation}
\end{equation}
where $\gamma_d$ is the pure dephasing rate. 
This model for dephasing is equivalent to the standard density matrix description where a decay rate $\gamma_d$ is added to the off-diagonal elements of the density matrix and has e.g. been shown to provide a good description for phonon induced dephasing in quantum dots \cite{Pertru,GoodPetru}.

We can now use the above model above to evaluate the desired correlation functions. From Eq. (\ref{eq:fi_fj}) we have
\begin{align}
\begin{split}
\expval{\vert{\alpha_{ij}}\vert^2} &= \int_{t_0}^{\infty}\int_{t_0}^{\infty} dtdt'\expval{f_i^{*}(t,t_0)f_j^{*}(t',t_0)f_i(t',t_0)f_j(t,t_0)}\,.
\label{eq:expval_alpha_1}
\end{split}
\end{align}
 Substituting Eq. \eqref{eq:languevin_correlation} into \ref{eq:expval_alpha_1}, assuming that the energy level in the quantum dot evolves slowly enough in time such that $\Delta_i = \Delta_j$, we find
\begin{align}
\begin{split}
\expval{\vert{\alpha_{ij}}\vert^2} = \iint \limits_{t_0}^{\infty}&dtdt' \gamma^2\exp{-\gamma(t+t'-2t_0)} \expval{\exp{-i\left(\phi_i(t') + \phi_j(t) - \phi_i(t) - \phi_j(t') \right)}}\,.
\end{split}
\label{eq:expval_alpha_2}
\end{align}
Since white noise is necessarily Gaussian, we have a Gaussian distribution for $\phi_i(t)$, leading to 
\begin{align}
\begin{split}
&\expval{\vert{\alpha_{ij}}\vert^2} = \iint \limits_{t_0}^{\infty}dtdt' \gamma^2\exp{-\gamma(t+t'-2t_0)} \exp{\expval{-\frac{\expval{\left(\phi_i(t') + \phi_j(t) - \phi_i(t) - \phi_j(t') \right)^2}}{2}}}\,.
\end{split}
\label{eq:expval_alpha_3}
\end{align}
 Applying Eq. \eqref{eq:languevin_correlation} the expectation value in the exponent yields
\begin{align}
\begin{split}
&\expval{\left(\phi_i(t') + \phi_j(t) - \phi_i(t) - \phi_j(t') \right)^2}  
= 4 \gamma_d |t'-t|\,,
\end{split}
\label{eq:expval_alpha_4}
\end{align}
finally after substituting in Eq. (\ref{eq:expval_alpha_3}), this leads to
\begin{equation}
\expval{\vert{\alpha_{ij}}\vert^2} = \frac{\gamma}{\gamma+2\gamma_d}\,.   
\end{equation}

In a similar fashion, one can obtain higher order terms, that are relevant for the three and four photon contributions:
\begin{align}
\begin{split}
\expval{\alpha_{ij}\alpha_{ik}\alpha_{jk}} &= \frac{\gamma^2}{\left(\gamma + \gamma_d\right)\left(\gamma + 2\gamma_d\right)}, \quad 
\expval{\alpha_{ij}\alpha_{kl}\alpha_{ik}\alpha_{jl}} = \frac{\gamma^3\left(3\gamma + 5\gamma_d\right)}{\left(\gamma + \gamma_d\right)\left(\gamma + 2\gamma_d\right)^2\left(3\gamma + 2\gamma_d\right)}\,,
\end{split}
\end{align}
where $i\neq j \neq k$. Note that $\expval{\alpha_{ij}\alpha_{kl}}=\expval{\alpha_{ij}}\expval{\alpha_{kl}}$ for $i,j\neq k,l$, since the noise is uncorrelated.

On the other hand, when the photons are generated at different stations, we can no longer assume that the energy splitting is zero during each run of the experiment. The slow drift implies a splitting $\Delta_{ij} \equiv \Delta_j - \Delta_i \neq 0 $, yielding the following expectation value of the indistinguishability parameter $\beta_{ij}$:
\begin{align}
\begin{split}
\expval{\vert{\beta_{ij}}\vert^2} =\iint \limits_{t_0}^{\infty}&dtdt' \gamma^2\exp{-\gamma(t+t'-2t_0)} \exp{-i\Delta_{ij}\left(t-t'\right)-2 \gamma_d |t'-t|}\,,
\end{split}
\label{eq:expval_beta_1}
\end{align}
where we have again calculated the expectation value as in Eq. (\ref{eq:expval_alpha_4}). Solving these integrals gives us 
\begin{equation}
\expval{\vert{\beta_{ij}}\vert^2} = \frac{\gamma\left(\gamma+2\gamma_d\right)}{\left(\gamma+2\gamma_d\right)^2+\Delta_{ij}^2}\,.   
\end{equation}

Similarly we can obtain the higher order terms that are needed for calculating the necessary correlations:
%\begin{widetext}
\begin{align}
\begin{split}
&\expval{\alpha_{ij}\beta_{ik}\beta_{jk}} = \frac{\gamma^2 \left(12 \left(\gamma +\gamma_d\right)^2 \left(\gamma + 2\gamma_d\right)^2 + \left(3\gamma^2 + 6\gamma\gamma_d + 4\gamma_d^2\right) \Delta^2\right)}{\left(3 \left(\gamma + \gamma_d\right) \left(\gamma + 2\gamma_d\right) \left(4 \left(\gamma +\gamma_d\right)^2 + \Delta^2\right) \left(\left(\gamma + 2\gamma_d\right)^2 + \Delta^2\right)\right) }\,, \\   \\
&\expval{\alpha_{ij}\alpha_{kl}\beta_{ik}\beta_{jl}} = \frac{\gamma}{3\gamma + 2\gamma_d}\left(-\frac{8\gamma_d \left(\gamma +\gamma_d\right) \left(2\gamma + \gamma_d\right)}{\left(\gamma + 2\gamma_d\right) \left(4 \left(\gamma +\gamma_d\right)^2 + \Delta^2\right)} + \frac{3\gamma^2 + 6\gamma\gamma_d + 2\gamma_d^2}{\left(\gamma + 2\gamma_d\right)^2 + \Delta^2} + \frac{2\gamma_d^2 \left(3\gamma + 2\gamma_d\right)^2}{\left(\gamma + 2\gamma_d\right)^2 \left(\left(3\gamma + 2\gamma_d\right)^2 + \Delta^2\right)}\right)\,.
\end{split}
\end{align}
%\end{widetext}
Note that as we assume that all photons are generated by only two quantum dots, there is in fact only one value of the detuning. Therefore we can reduce it to a single value $\Delta$ regardless of the considered index. 

To apply the above results we need to average over the slow variations of $\Delta$. We assume a Gaussian distribution centered around $\Delta=0$ with standard deviation $\sigma$ described by $P(\Delta)=1/\left(\sqrt{2\pi}\sigma\right)\exp{\frac{-\Delta^2}{2\sigma^2}}$. When we take the average we obtain:
%\begin{widetext}
\begin{align}
    \begin{split}
        \expval{\vert{\beta_{ij}}\vert^2} &= \sqrt{\frac{\pi}{2}}\frac{\gamma}{\sigma}e^{\frac{(\gamma+2\gamma_d)^2}{2\sigma^2}}\text{erfc}\left(\frac{\gamma+2\gamma_d}{\sqrt{2}\sigma}\right)\,, \\ 
        \expval{\alpha_{ij}\beta_{ik}\beta_{jk}} &= \sqrt{\frac{\pi}{2}} \frac{\gamma}{3\sigma(\gamma+\gamma_d)(\gamma+2\gamma_d)}\cdot \\
        &\left(e^{\frac{(\gamma+2\gamma_d)^2}{2\sigma^2}}(\gamma+2\gamma_d)(3\gamma+2\gamma_d)\text{erfc}\left(\frac{\gamma+2\gamma_d}{\sqrt{2}\sigma}\right) - 4e^{\frac{2(\gamma+\gamma_d)^2}{\sigma^2}}\gamma_d(\gamma+\gamma_d)\text{erfc}\left(\frac{\sqrt{2}(\gamma+\gamma_d)}{\sigma}\right)\right)\,, \\
        \expval{\alpha_{ij}\alpha_{kl}\beta_{ik}\beta_{jl}} &= \sqrt{\frac{\pi}{2}} \frac{\gamma}{\sigma(3\gamma+2\gamma_d)(\gamma+2\gamma_d)^2}(2e^{\frac{(3\gamma+2\gamma_d)^2}{2\sigma^2}}\gamma_d^2(3\gamma+2\gamma_d)\text{erfc}\left(\frac{3\gamma+2\gamma_d}{\sqrt{2}\sigma}\right)+\\
        e^{\frac{(\gamma+2\gamma_d)^2}{2\sigma^2}}&(\gamma+2\gamma_d)(3\gamma^2+6\gamma\gamma_d+2\gamma_d^2)\text{erfc}\left(\frac{\gamma+2\gamma_d}{\sqrt{2}\sigma}\right)- 4e^{\frac{2(\gamma+\gamma_d)^2}{\sigma^2}}\gamma_d(2\gamma+\gamma_d)(\gamma+2\gamma_d)\text{erfc}\left(\frac{\sqrt{2}(\gamma+\gamma_d)}{\sigma}\right))\,.
    \end{split}
\label{eq:full_ssigma_equations}
\end{align}
%\end{widetext}

Although exact, the above expressions give trouble numerically when $\sigma\rightarrow0$. Therefore, an asymptotic expansion of the complementary error functions has been used for $\sigma<0.1$. We express it up to third non-vanishing order
\begin{equation}
    \text{erfc}(x) \simeq \frac{1}{\sqrt{\pi}} e^{-x^2} \left(\frac{1}{x}-\frac{1}{2x^3}+\frac{3}{4x^5} \right)\,.
\end{equation}
For instance, $ \expval{\vert\beta_{ij}\vert^2}$ becomes
\begin{equation}
    \expval{\vert\beta_{ij}\vert^2} \simeq \frac{\gamma}{\gamma+2\gamma_d}\left(1 - \frac{\sigma^2}{(\gamma+2\gamma_d)^2} + \frac{3\sigma^4}{(\gamma+2\gamma_d)^4} \right)\,, 
\end{equation}
which shows that $ \expval{\vert\beta_{ij}\vert^2}(\sigma=0)= \expval{\vert\alpha_{ij}\vert^2}$, as expected. In a similar way, we calculate the expansion for higher order of the indistinguishability parameters.

\section{Photon operator transformations}\label{app:BS}

In this appendix we show explicitly how the optical set up transforms the initial creation operators and give more details of the model we use to describe the different errors mechanisms. From Eqs. \eqref{eq:BSrelations2} and \ref{eq:BSrelations3} we obtain the global transformation
\begin{align}
\begin{split}
\hat{a}_{s,{\varsigma}}^{\dagger}&=\frac{\sqrt{\eta_t T}}{2}\left(i\left(\hat{p}_H^{\dagger}\pm\hat{p}_V^{\dagger}\right)+\left(\hat{q}_H^{\dagger}\pm\hat{q}_V^{\dagger}\right)\right) +i\sqrt{1-T}\hat{a}_{\varsigma}^{\dagger}\,,
\end{split}
\label{eq:operator_relations}
\end{align} 
and similarly for Bob's photons, where $\varsigma\in\{H,V\}$.
We now apply Eq. \eqref{eq:operator_relations} to all operators in the initial state of the system $\hat{a}_{s,H}^{\dagger}\hat{a}_{s,V}^{\dagger}\hat{b}_{s,H}^{\dagger}\hat{b}_{s,V}^{\dagger}\ket{\emptyset}$ and trace out the CHS operators $\hat{p}_H^{\dagger}$, $\hat{p}_V^{\dagger}$, $\hat{q}_H^{\dagger}$ and $\hat{q}_V^{\dagger}$ to obtain the state shared by Alice and Bob after the heralding. This state, as explained in section \ref{sec:protocol}, will depend on the detection pattern at the CHS. If we ignore all imperfections and restrict the state to those events where photons lead to clicks corresponding to opposite polarizations 
\begin{align}
\begin{split}
\ket{\psi_{D1D4}} &= \frac { { \eta  }_{ t }{ T }{ (1-T) } }{ \sqrt{2} } \ket{\psi^-} = -\ket{\psi_{D2D3}}\,,
\end{split}
\label{eq:psi_14_23}
\end{align}
whereas the state that is created after the patterns $D_1D_2$ and $D_3D_4$ is 
\begin{equation}
\ket{\psi_{D1D2}} = \ket{\psi_{D3D4}} = \frac{i{ \eta  }_{ t }{ T }{ (1-T) }}{\sqrt{2}} \ket{\phi^-}\,,
\label{eq:psi_12_34}
\end{equation}
agreeing with our HOM arguments from section \ref{sec:protocol}.

In this work we focus the analysis on the patterns that generate the state $\ket{\psi^{-}}$ ($D_1D_4$ and $D_2D_3$) in order to simplify the presentation of the results. Note that the transmission probability only enters in the prefactor of the unnormalised state. After renormalization to account for the state conditioned on the detection of the photons, the states are independent of the transmission efficiency thanks to the heralding scheme.

The heralding station notifies Alice and Bob when the correct pattern of clicks has been detected. In the ideal scenario these patterns guarantee that the protocol succeeded in generating the desired state. To evaluate the performance under non-ideal condition, however, we need to consider the error mechanisms described in the previous section. To do this we start by describing the transformation of the mode operators when a photon is lost:
\begin{equation}
\hat{a}^{\dagger}_{i} \rightarrow \sqrt{\eta_{j,i}} \hat{a}^{\dagger}_{i} + \sqrt{1-\eta_{j,i}} \hat{c}_{j,i}^{\dagger}\,,
\label{eq:transf_loss}
\end{equation}
where $\hat{c}_{j,i}^{\dagger}$ creates a photon that escapes the set up. The index $j \in\{ 1,2,t\}$ indicates where the loss happened following the notation from Fig. \ref{fig:losses}. Applying this transformation to each segment of the setup leads us to the loss operator $\hat{L}_{i}^{\dagger}$:
\begin{align}
\begin{split}
\hat{L}_{i}^{\dagger} \equiv \sqrt{1-\eta_{1,i}} \hat{c}_{1,i}^{\dagger} +  i\sqrt{\eta_{1,i}(1-\eta_{2,i})(1-T)} \hat{c}_{2,i}^{\dagger}  +  \sqrt{\eta_{1,i}(1-\eta_t)T} \hat{c}_{t,i}^{\dagger} \,.
\end{split}
\label{eq:f_i_operator_lost}
\end{align} 
The operator directly provides the total probability for the photon to escape the set-up $\expval{\hat{L}_{i}\hat{L}_{i}^{\dagger}} =  1-\eta_{1,i}\eta_{2,i}(1-T) - \eta_{1,i}\eta_{t}T$. Note that $\expval{\hat{L}_{i}\hat{L}_{j}^{\dagger}}_{i\neq j} = 0$ and $\expval{\hat{L}_{i}\hat{L}_{j}\hat{L}_{k}^{\dagger}\hat{L}_{l}^{\dagger}}_{i\neq k, j\neq l} = 0$ expressing that lost photons do not interfere. By means of Eqs. \eqref{eq:transf_loss} and \ref{eq:f_i_operator_lost} we can obtain the global transformation in the main text (\ref{eq:f_i_operator_relations}).

Finally, we detail the action of the consecutive QWP and HWP on the creation operators $\hat{a}_{H,{1}}^{\dagger}$, $\hat{a}_{V,{2}}^{\dagger}$, $\hat{b}_{H,{3}}^{\dagger}$ and $\hat{b}_{V,4}^{\dagger}$, represented by the Jones matrices. They are described by the unitary operations \cite{HWP}:
\begin{align}
\begin{split}
\hat{U}_{QWP}(\theta) &= \frac{1}{\sqrt{2}}  \begin{pmatrix} i - \cos(2\theta) & \sin(2\theta) \\ \sin(2\theta) & i+\cos(2\theta) \end{pmatrix}, \quad 
\hat{U}_{HWP}(\phi) = \begin{pmatrix} \cos(2\phi) & -\sin(2\phi) \\ -\sin(2\phi) & -\cos(2\phi) \end{pmatrix}\,.
\end{split}
\end{align}
The total transformation is thus $\hat{U}_{tot}(\theta,\phi)= \hat{U}_{HWP}(\phi)\hat{U}_{QWP}(\theta)$.

\section{Conditioning on the correct detection pattern}\label{app:O}

In this appendix we explicitly show the expectation values of the CHS operators $\hat{O}_{i}$ up to three-photon detections as a function of the indistinguishability parameters $\alpha_{ij}$ and $\beta_{ij}$. Selecting only the combinations of operators that lead to a correct event, that is, $D_1D_4$ and $D_2D_3$ and applying Eqs. \eqref{eq:fi_fj} and (\ref{eq:O_operators}), we obtain: 
\begin{align}
\begin{split}
&\expval{\hat{O}_{1}\hat{O}_{3}\hat{O}_{1}^{\dagger}\hat{O}_{3}^{\dagger}}_{D1D4,D2D3} = \\
&= \frac{1}{2}\expval{\left(\left( \hat{p}_{H,{3}}\hat{q}_{H,{1}} + \hat{p}_{V,{3}}\hat{q}_{H,{1}} \right) - \left(  \hat{p}_{H,{1}}\hat{q}_{H,{3}} + \hat{p}_{V,{1}}\hat{q}_{H,{3}} \right)  \right) \cdot (( \hat{p}_{H,{3}}^{\dagger}\hat{q}_{H,{1}}^{\dagger} + \hat{p}_{V,{3}}^{\dagger}\hat{q}_{H,{1}}^{\dagger} ) - (  \hat{p}_{H,{1}}^{\dagger}\hat{q}_{H,{3}}^{\dagger} + \hat{p}_{V,{1}}^{\dagger}\hat{q}_{H,{3}}^{\dagger} ) )} = 1 - \vert \beta_{13} \vert ^2 \,.
\end{split}
\end{align}
Similarly, for the other combinations:
\begin{align}
\begin{split}
\expval{\hat{O}_{1}\hat{O}_{4}\hat{O}_{1}^{\dagger}\hat{O}_{4}^{\dagger}}_{D1D4,D2D3} &= 2t(1-t)1 + \vert\beta_{14}\vert^2, \quad 
\expval{\hat{O}_{2}\hat{O}_{3}\hat{O}_{2}^{\dagger}\hat{O}_{3}^{\dagger}}_{D1D4,D2D3} = 1 + \vert\beta_{23}\vert^2, \\ 
\expval{\hat{O}_{2}\hat{O}_{4}\hat{O}_{2}^{\dagger}\hat{O}_{4}^{\dagger}}_{D1D4,D2D3} &= 1 - \vert\beta_{24}\vert^2, \quad 
\expval{\hat{O}_{1}\hat{O}_{2}\hat{O}_{1}^{\dagger}\hat{O}_{2}^{\dagger}}_{D1D4,D2D3} = 1 - \vert\alpha_{12}\vert^2, \\
\expval{\hat{O}_{3}\hat{O}_{4}\hat{O}_{3}^{\dagger}\hat{O}_{4}^{\dagger}}_{D1D4,D2D3} &=  1 - \vert\alpha_{34}\vert^2, \quad
 \expval{\hat{O}_{1}\hat{O}_{4}\hat{O}_{2}^{\dagger}\hat{O}_{3}^{\dagger}}_{D1D4,D2D3} = - \alpha_{12}\alpha_{34} - \beta_{13}\beta_{24}, \\  \expval{\hat{O}_{1}\hat{O}_{3}\hat{O}_{2}^{\dagger}\hat{O}_{4}^{\dagger}}_{D1D4,D2D3} &=  - \alpha_{12}\alpha_{34} + \beta_{14}\beta_{23}\,.
 \label{eq:2exp_val}
\end{split}
\end{align}
It can be easily checked that any other combination of CHS operators has vanishing expectation value. If no conditioning on the detection pattern is done, the sum of the expectation values are independent of the indistinguishability of the photons and equal 4, reflecting that with the current normalization this is proportional to the product of the number of photons at each side. Note that in the limit of completely indistinguishable photons ($\alpha_{12}=\alpha_{34}=1$), $\expval{\hat{O}_{1}\hat{O}_{2}\hat{O}_{1}^{\dagger}\hat{O}_{2}^{\dagger}}_{D1D4,D2D3}$ and $\expval{\hat{O}_{3}\hat{O}_{4}\hat{O}_{3}^{\dagger}\hat{O}_{4}^{\dagger}}_{D1D4,D2D3}$ vanish, 
meaning that we never have a successful detection event with two photons from the same station. As explained in Sec. \ref{sec:protocol}, this occurs because of the HOM effect, since two photons from from the same station will bunch together after the half-wave plate ($\hat{a}_{H,{1}}^{\dagger}\hat{a}_{V,{2}}^{\dagger} \rightarrow \hat{a}_{H}^{'\dagger 2} - \hat{a}_{V}^{'\dagger 2}  $), and therefore no $D_1D_4$ or $D_2D_3$ clicks can ever happen. Finally, note that the expectation values in Eq. \eqref{eq:2exp_val} correspond to the generalised matrix elements of the density matrix in Eq. \eqref{eq:rho_tr_O_i}.

In a similar manner, we can calculate the expectation values for the occurrence of three photons at the CHS. We only keep those terms that correspond to combinations of clicks perceived as correct by the CHS, $D_1D_4$ and $D_2D_3$, due to the detectors not being photon-number resolving. For instance, a term such $\hat{p}_{H,{1}}^{\dagger}\hat{q}_{V,{3}}^{\dagger}\hat{q}_{V,{4}}^{\dagger}$ is taken into consideration, while $\hat{p}_{H,{1}}^{\dagger}\hat{p}_{V,{3}}^{\dagger}\hat{q}_{V,{4}}^{\dagger}$ is not. Applying once more Eqs. \eqref{eq:fi_fj} and (\ref{eq:O_operators}) we obtain
\begin{align}
\begin{split}
\expval{\hat{O}_{1}\hat{O}_{3}\hat{O}_{4}\hat{O}_{1}^{\dagger}\hat{O}_{3}^{\dagger}\hat{O}_{4}^{\dagger}}_{D1D4,D2D3} &= \frac{1}{2}\left(
3-\abs{\alpha_{34}}^2 -\abs{\beta_{13}}^2 + 3\abs{\beta_{14}}^2  -2\alpha_{34}\beta_{13}\beta_{14} \right), \\
\expval{\hat{O}_{2}\hat{O}_{3}\hat{O}_{4}\hat{O}_{2}^{\dagger}\hat{O}_{3}^{\dagger}\hat{O}_{4}^{\dagger}}_{D1D4,D2D3} &= \frac{1}{2}\left(
3-\abs{\alpha_{34}}^2 +3\abs{\beta_{23}}^2 - \abs{\beta_{24}}^2  -2\alpha_{34}\beta_{23}\beta_{24} \right), \\
\expval{\hat{O}_{3}\hat{O}_{1}\hat{O}_{2}\hat{O}_{3}^{\dagger}\hat{O}_{1}^{\dagger}\hat{O}_{2}^{\dagger}}_{D1D4,D2D3} &= \frac{1}{2}\left(
3-\abs{\alpha_{12}}^2 -\abs{\beta_{13}}^2 + 3\abs{\beta_{23}}^2  -2\alpha_{12}\beta_{13}\beta_{23} \right), \\
\expval{\hat{O}_{4}\hat{O}_{1}\hat{O}_{2}\hat{O}_{4}^{\dagger}\hat{O}_{1}^{\dagger}\hat{O}_{2}^{\dagger}}_{D1D4,D2D3} &= \frac{1}{2}\left(
3-\abs{\alpha_{12}}^2 +3\abs{\beta_{14}}^2 - \abs{\beta_{24}}^2  -2\alpha_{12}\beta_{14}\beta_{24} \right)\,, 
\end{split}
\end{align}
where again any other combination of three CHS operators has an expectation value equal zero.

\section{Probability events $P_{ijkl}$}\label{app:table_probabilities}

In this section we include Table \ref{tab:bigtable}, where we detail the contribution of all 4 and 5-photon events.
\vspace{1cm}
\twocolumngrid

\begin{table}
\begin{ruledtabular}
\begin{tabular}{lllll}
Event probability   & CHS & Alice       & Bob         & Lost photons \\ \hline \hline
$P_{2110}$ & AB        & A           & B & $\emptyset$  \\ \hline
$P_{2200}$ & BB        & AA           & $\emptyset$ & $\emptyset$  \\ \hline
$P_{2020}$ & AA        & $\emptyset$           & BB & $\emptyset$  \\ \hline
$P_{2101}$ & AB         & A           & $\emptyset$ & B            \\
           & BB         & A           & $\emptyset$ & A            \\ \hline
$P_{2011}$ & AB         & $\emptyset$ & B           & A            \\
           & AA         & $\emptyset$ & B           & B            \\ \hline
$P_{2002}$ & AB         & $\emptyset$ & $\emptyset$ & AB           \\
           & AA         & $\emptyset$ & $\emptyset$ & BB           \\
           & BB         & $\emptyset$ & $\emptyset$ & AA           \\ \hline
$P_{3100}$ & ABB        & A           & $\emptyset$ & $\emptyset$  \\ \hline
$P_{3010}$ & AAB         & $\emptyset$ & B           & $\emptyset$  \\ \hline
$P_{3001}$ & ABB        & $\emptyset$ & $\emptyset$ & A            \\
           & AAB        & $\emptyset$ & $\emptyset$ & B            \\ \hline \hline
$P_{2111}$ & AB        & A & B & A            \\
           & AA        & A & B & B            \\ \hline
$P_{2210}$ & AB        & AA           & B & $\emptyset$  \\ \hline
$P_{2120}$ & AA        & A           & BB & $\emptyset$  \\ \hline  
$P_{2201}$ & AB        & AA & $\emptyset$ & B            \\
           & BB        & AA & $\emptyset$ & A            \\ \hline
$P_{2021}$ & AA       & $\emptyset$ & BB & A            \\\hline
$P_{2102}$ & AB        & A & $\emptyset$ & AB            \\
           & AA        & A & $\emptyset$ & BB            \\ 
           & BB        & A & $\emptyset$ & AA            \\\hline
$P_{2012}$ & AB        & $\emptyset$ & B & AA            \\
           & AA        & $\emptyset$ & B & AB             \\ \hline
$P_{2300}$ & BB        & AAA           & $\emptyset$ & $\emptyset$  \\ \hline
$P_{2030}$ & AA        & BBB           & $\emptyset$ & $\emptyset$  \\ \hline
$P_{2003}$ & AB        & $\emptyset$     & $\emptyset$ & AAB  \\
	       & AA        & $\emptyset$     & $\emptyset$ & ABB  \\
		   & BB        & $\emptyset$     & $\emptyset$ & AAA  \\ \hline
$P_{3110}$ & AAB        & A           & B & $\emptyset$  \\ \hline
$P_{3101}$ & AAB        & A & $\emptyset$  & B            \\
           & ABB        & A & $\emptyset$ & A            \\ \hline
$P_{3011}$ & AAA        & $\emptyset$ & B  & B            \\
           & AAB        & $\emptyset$ & B & A            \\ \hline
$P_{3200}$ & ABB        & AA           & $\emptyset$ & $\emptyset$  \\ \hline
$P_{3020}$ & AAA        & $\emptyset$           & BB & $\emptyset$  \\ \hline
$P_{3002}$ & AAA        & $\emptyset$ & $\emptyset$ & BB            \\
           & AAB        & $\emptyset$ & $\emptyset$ & AB            \\ 
           & ABB        & $\emptyset$ & $\emptyset$ & AA \\       
\end{tabular}
\end{ruledtabular}
\caption{\label{tab:events} 4-photon and 5-photon events that lead to a correct detection pattern at the CHS. A and B indicate where the photons were generated, while the header tells where they were detected.}
\label{tab:bigtable}
\end{table}

%\twocolumngrid
%\nocite{*}

\bibliography{mybib_revtex}% Produces the bibliography via BibTeX.

\end{document}